\documentclass[aps,prl,notitlepage,nofootinbib]{revtex4-1}
\usepackage[a4paper,top=1.25in,bottom=1.25in,left=1.250in,right=1.250in]{geometry}
\usepackage{graphicx}
\usepackage{color}
\usepackage{ifpdf}
\usepackage{epstopdf}
\DeclareGraphicsExtensions{.eps, .pdf, .jpg, .tif}
\usepackage[dvips]{hyperref}

\newcommand{\be}{\begin{equation}}
\newcommand{\ee}{\end{equation}}
\newcommand{\ber}{\begin{eqnarray}}
\newcommand{\eer}{\end{eqnarray}}
\newcommand{\qar}{\begin{array}{l}}
\newcommand{\ear}{\end{array}}
\def\mp{{\hspace*{-1pt}+\hspace*{-1pt}}}
\def\mm{{\hspace*{-1pt}-\hspace*{-1pt}}}
\def\mdot{{\hspace*{-1pt}\cdot\hspace*{-1pt}}}
\begin{document}
\title{
\vspace*{-2cm}\hspace*{-2cm}
{\tiny
\emph{Superconductivity: From Basic Physics to the Latest Developments}, World Scientific: 45-78 (1995).
\href{http://dx.doi.org/10.1142/9789814503891_0002}{doi:10.1142/9789814503891\_0002}
}
\\
\vspace*{2cm}
{Strong Coupling Theory of Superconductivity}
}
\author{D. Rainer}
\affiliation{
Physikalisches Institut, Universit\"at 
Bayreuth, D-95440 Bayreuth, Germany
}
\author{J. A. Sauls}
\affiliation{
Nordita, Blegdamsvej 17, DK-2100 Copenhagen, {\O} Denmark \\
Department of Physics \& Astronomy, Northwestern University, 
Evanston, IL 60208 USA 
}
\date{October 15, 1992}
\begin{abstract}
The electronic properties of correlated metals with  a strong
electron-phonon coupling may be understood in terms of a combination of
Landau's Fermi-liquid theory and the strong-coupling theory of Migdal
and Eliashberg.  In these lecture notes we discuss the microscopic
foundations of this phenomenological Fermi-liquid model of correlated,
strong-coupling metals. We formulate the basic equations of the model,
which are quasiclassical transport equations that describe both
equilibrium and non-equilibrium phenomena for the normal and
superconducting states of a metal. Our emphasis is on superconductors
close to equilibrium, for which we derive the general linear response
theory. As an application we calculate the dynamical conductivity of
strong-coupling superconductors.
\end{abstract}
\maketitle
\bigskip
\section{Introduction}

\noindent Traditionally, ``strong-coupling superconductivity'' is
understood as superconductivity induced by a very strong
phonon-mediated pairing interaction. The strong-coupling theory was
developed by Eliashberg\cite{eli60} on the basis of Migdal's theory for
the normal state \cite{mig58}. In strong-coupling superconductors the
transition temperature $T_c$ is comparable to the Debye temperature.
Thus, typical energies of the superconducting electrons, e.g. $k_BT$ or
the superconducting energy gap, are comparable to typical energies of
the excitations mediating the attractive interaction.  In this case the
superconductivity is sensitive to the dynamic properties of these
excitations, which leads to deviations from the universal laws of
weak-coupling superconductivity (law of corresponding states) and, more
importantly, opens the possibility of extracting information on the
origin of the attractive interaction from superconducting measurements.
The most prominent example is McMillan's \cite{mcm65,mcm69}  analysis
of superconducting tunneling data  which gives us the spectral function
$\alpha^2F(\omega)$ of the phonon-mediated attractive interaction
together with the  instantaneous electronic interaction parameter
$\mu^{\ast}$.\cite{bog58,mor62} The electronic part of the pairing
interaction is found to be repulsive (positive $\mu^{\ast}$) in all
cases where it has been measured.  However, attractive electronic
interaction mechanisms, i.e. negative $\mu^{\ast}$'s, are conceivable
and have been suggested as a possible origin of high transition
temperatures.\cite{lit64,gin64,gin82} The material parameters of the
strong-coupling theory, $\alpha^2F(\omega)$ and $\mu^{\ast}$, have been
obtained from  tunneling experiments for several superconductors. These
material parameters can be used as input into the strong-coupling
theory to calculate numerically strong-coupling corrections to the
universal weak-coupling results.  Such calculations were first
performed by Scalapino, Wada, Swihart, Schrieffer, and
Wilkins\cite{sca65,sca66} and have since been developed to some degree
of perfection. Efficient computer codes have been designed for
calculating $T_c$, the energy gap, the isotope effect, critical fields,
magnetization curves, the vortex lattice, the electromagnetic response,
and other measurable quantities of interest. The basic equations,
numerical methods, and various results are documented in several
textbooks \cite{sch64,gri81,all82,gin82,von82} and review articles
\cite{sca69,mcm69,all80,dol82,rai86,car90}. In table
\ref{tab_delta} we compare calculated and measured results for the ratio
$2\Delta_0/k_BT_c$, which has the universal value 3.53 for isotropic
weak-coupling superconductors. There is generally very good agreement
between the calculated strong-coupling corrections and measured data,
which demonstrates the high accuracy of Eliashberg's strong-coupling
theory for traditional superconductors. A recent thorough discussion of 
strong-coupling effects in the ratio 
$2\Delta_0/k_BT_c$ can be found in Ref. \cite{com93}. One might say that
strong-coupling superconductivity is the best understood many-body
effect.

\par

\begin{table}
\begin{center}
\label{tab_delta}
\begin{tabular}{|c|c|c|c|}
\hline
 Metal &Tunneling &Weak-coupling &Strong-coupling \\
  &experiment &theory (error) & theory (error)\\
\hline 
Hg &4.61 & 3.53 (+31\%) & 4.50 (+ 2\%)\\
In &3.69 & 3.53 (+ 5\%) & 3.87 (-- 5\%)\\
Nb &3.80 & 3.53 (+ 8\%) & 3.83 (-- 1\%)\\
Pb &4.50 & 3.53 (+27\%) & 4.47 (+ 1\%)\\
Sn &3.74 & 3.53 (+ 6\%) & 3.73 (+ 0\%)\\
Tl &3.60 & 3.53 (+ 2\%) & 3.71 (-- 3\%)\\
\hline 
\end{tabular}
\end{center}
\caption{Measured and calculated ratios $2\Delta_0/k_BT_c$. The errors indicate the 
necessary corrections required to obtain the
experimental result.}
\end{table}
\par
\bigskip

In these lecture notes we focus on specific aspects of the
strong-coupling theory, all of which are connected with the problem of
superconductivity in strongly correlated electron systems. We consider
Eliashberg's strong-coupling theory as a generalization of the
Fermi-liquid theory of superconductivity to metals with strong
electron-phonon coupling. Like Landau's Fermi-liquid theory it lives on
the border line between microscopic theories and phenomenological
theories. Landau's and Eliashberg's theories can be derived
microscopically by asymptotic expansions in small parameters such as
$k_BT/E_F$, $\hbar\omega_D/E_F$, etc. On the other hand, they contain
material parameters (Landau parameters, $\alpha^2F(\omega)$,
$\mu^{\ast}$, and others) which are phenomenological parameters to be
taken from experiments. These material parameters have a precise
microscopic meaning, but their calculation from first principles seems
outside the reach of
present many-body techniques.

\par

In the first part of this lecture (Sections II and III) we present the conceptual background of
the strong-coupling theory. We discuss the derivation of the
theory by asymptotic expansions, specify the various material parameters,
and formulate the strong-coupling theory in terms of a quasiclassical
transport equation. This equation is the generalization of the
Boltzmann-Landau transport equation for normal Fermi liquids to  
superconductors with strong electron-phonon coupling. We keep the 
formulation general enough to cover metals with strongly 
anisotropic Fermi surfaces. 
\par

In the second part (Section IV) we discuss formal aspects of the linear response
theory of Fermi-liquid superconductors with strong electron-phonon
coupling. We solve the quasiclassical transport equations to linear
order in the external perturbation, and derive the fundamental integral
equations for linear response coefficients. As a specific example, these equations 
are used to obtain the general formula for the electromagnetic response of 
a strong-coupling superconductor with impurity scattering. 

\section{Fermi-Liquid Model}

The strong-coupling theory is an accurate theory of superconductivity
which provides a quantitative explanation of essentially all
superconducting phenomena, including the observed deviations from the
universal laws of weak-coupling BCS theory.\cite{bcs57} On the other hand, this
theory was of no help in an important field of experimental
superconductivity, the search for new materials with higher $T_c$'s.
This part of the lecture deals basically with this apparent
discrepancy. By examining the roots of strong-coupling theory we can
understand  both a) the accuracy of the theory for  weakly and
strongly correlated systems, and b) the failure of the theory in
predicting $T_c$'s of new materials.\par

The strong-coupling theory is correct to leading orders in certain
expansion parameters, such as the ratios $k_BT_c/E_f$,
$\hbar\omega_D/E_f$, $1/k_f\ell$. The Fermi energy, $E_f$, is
representative of typical electronic energies (band width, interaction
energies, Hubbard U's, etc.), $k_BT_c$ is the characteristic energy
scale of the superconducting state (thermal energy, energy gap),
$\hbar\omega_D$ is a typical phonon energy, $1/k_f$ represents
microscopic length scales (Fermi wavelength, lattice constants,
screening length, etc.), and $\ell$ is the electron mean free path. In
standard metals we find superconducting energies of the order
10$^{-4}$--10$^{-2}$eV, electronic energies of $\approx$ 1--10eV,
phononic energies of $\approx$ 10$^{-2}$--10$^{-1}$eV, microscopic
length scales of $\approx$ 1{\AA}, and the mean free path of $\approx$
10--10$^4${\AA}. Hence, the expansion parameters of the
strong-coupling theory are of the order 10$^{-4}$--10$^{-1}$. We follow
our previous notations\cite{ser83,rai86} and assign the expansion
parameters the order of magnitude {\small {\sl small}}. A
perturbation expansion in powers of {\small {\sl small}} cannot be done
with standard quantum mechanical perturbation methods, because {\small
{\sl small}} does not refer, in general, to small terms in the
Hamiltonian. The problem of an asymptotic expansion in {\small {\sl
small}} was basically solved by Landau for strongly interacting
electrons in their normal state,\cite{lan59,eli62} by Migdal\cite{mig58} for normal metals with
strong electron-phonon coupling, by Abrikosov and Gorkov for dirty
metals,\cite{abr59,agd63} and by Eliashberg\cite{eli60} for superconducting metals with strong
electron-phonon coupling. The model of a metal in its normal and
superconducting states, which comprises all these theories will be
called the ``Fermi-liquid model'' in the following. All microscopic
derivations of the Fermi-liquid model start from a formulation of the
many-body problem in terms of many-body Green's functions.
\cite{agd63,fet71,ric80}
 We also
use this traditional tool of many-body calculations in this section.
Following Landau we then transform the Green's function
formulation of the Fermi liquid-model into a quasiclassical transport
theory in the next section.
\par

\subsection*{Dyson's Equations}

The Green's function method\cite{agd63,fet71,ric80} utilizes several types of many-body Green's
functions. Each of them is optimal for special tasks. The Matsubara
Green's functions for electrons ($G^M$) and
            phonons (${\cal D}^M$), for
instance, are most convenient for calculating properties of metals in
thermal equilibrium. A set of three Green's functions, $G^R$
(``retarded''), $G^A$ (``advanced''), $G^K$ (``Keldysh'') for
electrons, and similarly, three additional Green's functions for the
    phonons, ${\cal D}^{R,A,K}$, 
are used in the Keldysh technique,\cite{kel65}
 which is the
most powerful method for non-equilibrium phenomena. We use  Keldysh's 
method in this lecture because it gives us directly the non-equilibrium
properties of superconductors, such as dynamical response functions,
relaxation phenomena, tunneling conductivities, etc., which are the
most important phenomena for measuring  strong-coupling effects.
The strong-coupling model in the Matsubara representation is well
documented in textbooks and reviews.\cite{all82,car90} We emphasize, that
we do not loose generality by favouring a special Green's function
technique. All techniques are formally equivalent, and the results
obtained in one technique, like the expansion in {\small {\sl small}}
or the derivation of a transport theory hold for the other techniques.
\par

The following discussion is based on the jellium model for a metal;\cite{agd63,fet71,ric80}
i.e. we ignore the lattice structure and assume the electrons to form a
liquid of negatively charged fermions coupled to a continuous, elastic
background of positive charges. The jellium model is a sensible model
for qualitative discussions. Realistic calculations take the lattice
structure into account, which requires a more involved `book-keeping'
scheme\cite{rai86}, more involved numerical work, but does not require
new concepts. We will return to a realistic, material-oriented model
in sections III and IV. 
In the standard space-time representation, the Green's
functions of the jellium model depend on two positions and two times,
i.e. $G^{R,A,K}\rightarrow G^{R,A,K}(\vec x_{1},\vec
x_{2};t_{1},t_{2})$, and 
                 ${\cal D}^{R,A,K} \rightarrow {\cal D}^{R,A,K}(\vec
R_{1},\vec R_{2};t_{1},t_{2})$. Equivalently, one might use the Fourier
transformed Green's functions, $G^{R,A,K} \rightarrow G^{R,A,K}(\vec
k_{1},\vec k_{2};\epsilon_1,\epsilon_2)$, 
                  and ${\cal D}^{R,A,K} \rightarrow {\cal D}^{R,A,K}(\vec q_{1},\vec q_{2};
\omega_{1},\omega_{2})$. Another choice
of variables, which is exact and more convenient for our purposes, is
the mixed representation, $G^{R,A,K} \rightarrow G^{R,A,K}(\vec p,\vec
               R;\epsilon,t)$, ${\cal D}^{R,A,K} \rightarrow {\cal D}^{R,A,K}(\vec q,\vec
R;\omega,t)$. This representation is optimally suited for a derivation
of the Fermi-liquid theory of metals. One can interpret $\vec p$
(=$(\vec k_1+\vec k_2)/2$) as the momentum of a single-electron
excitation, $\vec R$ (=$(\vec x_1+\vec x_2)/2$) as its location,
$\epsilon$ (=$(\epsilon_1+ \epsilon_2)/2$) as its energy, and $t$
(=$(t_1+t_2)/2$) as the time of the observer. Similarly, $\vec q$ and
$\omega$ are the average wave vector and frequency of an elastic
deformation of the positive background at position $\vec R$ and time
$t$. 

\par

The equations of motion for the Green's functions are Dyson's equations,
which in the mixed representation have the form
\be\label{dyson}
\left(\epsilon\hat\tau_3-\xi_0(\vec p)-\hat V(\vec p,\vec R;t)-
\hat\Sigma^{R,A}(\vec p,\vec R;\epsilon,t)\right)\otimes\hat G^{R,A}
(\vec p,\vec R;\epsilon,t)=\hat 1 \ ,
\ee
\be
\begin{array}{c}\label{dysonk}
\left(\epsilon\hat\tau_3-\xi_0(\vec p)-\hat V(\vec p,\vec R;t)-
\hat\Sigma^{R}(\vec p,\vec R;\epsilon,t)\right)\otimes\hat G^{K}
(\vec p,\vec R;\epsilon,t) 
-\\
\hat\Sigma^{K}(\vec p,\vec R;\epsilon,t)\otimes\hat G^{A}
(\vec p,\vec R;\epsilon,t)=0\ .
\end{array}
\ee
\be\label{dysonk1}
\left(-M\omega^2+\Pi^{R,A}(\vec q,\vec R;\omega,t)\right)\otimes {\cal D}^{R,A}
(\vec q,\vec R;\omega,t) =1\ , 
\ee
\be\label{dysonk2}
\begin{array}{l}
\left(-M\omega^2+\Pi^R(\vec q,\vec R;\omega,t)\right)\otimes {\cal D}^{K}
(\vec q,\vec R;\omega,t) +
\Pi^K(\vec q,\vec R;\omega,t)\otimes {\cal D}^{A}
(\vec q,\vec R;\omega,t)=0\ .
\end{array}
\ee

In this representation, the Keldysh Green's functions are  directly
related to the Wigner distribution function for many-particle systems.
The `hat' on the electron Green's functions $\hat G$, the external perturbation
$\hat V$, and the electron self-energies $\hat\Sigma$ indicates their
4$\times$4 Nambu-matrix structure, which comprises the spin and
particle-hole degrees of freedom. The electron kinetic
energy minus the chemical potential is denoted by  $\xi_0(\vec p)$, 
$\hat\tau_3$ is the third Pauli
matrix in particle-hole space, and $\hat 1$ is the 4-dimensional
unit-matrix. The ionic mass is denoted by M, and the 
phonon self-energies  by $\Pi$; they include
the elastic interactions of the positive background. The
$\otimes$-products are  defined as
\be\label{product}
\begin{array}{l}
\hspace{-1.0em}
[\hat A\otimes 
\hat B](\vec p,\vec R;\epsilon,t)=
e^{{i\over 2}\left( -\vec\nabla^A_R\cdot\vec\nabla^B_p
+\vec\nabla^A_p\cdot\vec\nabla^B_R +\partial^A_{\epsilon}\partial^B_t
-\partial^A_{t}\partial^B_{\epsilon}\right)}
\hat A(\vec p,\vec R;\epsilon,t)\hat B(\vec p,\vec R;\epsilon,t)\,,
\end{array}
\ee
\be\label{product1}
\begin{array}{l}
\hspace{-1.0em}
[A\otimes B](\vec q,\vec R;\omega,t)= 
e^{{i\over 2}\left( -\vec\nabla^A_R\cdot\vec\nabla^B_q
+\vec\nabla^A_q\cdot\vec\nabla^B_R +\partial^A_{\omega}\partial^B_t
-\partial^A_{t}\partial^B_{\epsilon}\right)}
A(\vec q,\vec R;\omega,t) B(\vec q,\vec R;\omega,t)\,,
\end{array}
\ee
where the superscripts $A$ ($B$) on 
the partial derivatives indicate 
differentiation with respect to the arguments of $A$ ($B$).
Note that we could equally well have chosen to write Dyson's equations with the 
inverse operators $(\epsilon\hat\tau_3-\xi_0-\hat V-
\hat\Sigma^{R,A})$, etc. on the right.

\subsection*{Skeleton Diagrams}

In Dyson's equations all difficult many-body physics is moved into the
self-energies $\Sigma$ and $\Pi$. The Fermi-liquid model for
superconductors with strong electron-phonon coupling amounts to a
specific, well founded approximation for the self-energies; one takes
into account all self-energy processes to zeroth- and first-order in
the expansion parameter {\small {\sl small}}. A convenient starting
point for obtaining the expansion in {\small {\sl small}} is the
standard decomposition of the self-energies into a sum over Feynman
diagrams. The diagrams are generated in the usual way by a formal
perturbation expansion in the electron-electron interaction, the
electron-phonon interaction, the electron-impurity interaction, the
phonon-phonon interaction, and the external potentials. Elements of
these diagrams are bare non-interacting Green's functions and bare
interaction vertices. The procedure is to classify each diagram according
to its order in {\small {\sl small}}, and to sum  the contributions
of all leading order diagrams. The following steps lead to a
classification of diagrams in the parameter {\small {\sl small}}.

\begin{enumerate}
\item The first step is to
formally split the
electron and phonon
Green's functions into a high-energy and low-energy part, $\hat G=\hat
G_h+\hat G_l$, and $D=D_h+D_l$. The low-energy part, $\hat G_l$, is
chosen to agree with the full $\hat G$ for energies near zero
($\mid\epsilon\mid < E_c$) and momenta near the Fermi surface $\mid p -p_f\mid
< k_c$, where $E_c$ and $k_c$ are technical cutoffs defining the
``low-energy range'' in energy-momentum space. $\hat G_l$ is zero outside
this low-energy range in energy-momentum space. The high-energy part, 
$\hat G_h$, agrees with $\hat G$ outside the low-energy range, but vanishes
inside $E_c$ and $k_c$. The separation of the phonon Green's function
is done in the same way, except for a modified definition of its
low-energy range. The low-energy range for phonons covers the frequency
range $\omega < E_c$ and {\it all} phonon-wave vectors $q$. The
technical cut-offs $E_c$ and $k_c$ are to be chosen in the window
between small and large energies and momenta 
({\small {\sl small}} $< E_c/E_f <1$, {\small {\sl small}} $< k_c/k_f <1$).
This separation into low- and high-energy
parts can also be expressed in terms of diagrams (see
\cite{ser83,rai86}). The new set of  diagrams is  
substantially enlarged, since any
traditional diagram with $n$ Green's
function lines generates $2^n$ new diagrams. An instructive 
example is shown in 
Figure 1.
 The advantage of the new diagrams is that they 
can be classified systematically according to
their order in {\small {\sl small}}. The order 
is determined by the number of 
low-energy Green's functions (thick lines), and how they are 
linked via interaction
vertices (open circles) and high-energy Green's 
functions (dashed lines). The diagram  on the left in 
Figure 1 
has four low-energy Green's functions, two high-energy
Green's functions, two electron-phonon vertices and two electron-electron
interaction vertices. We call  high-energy Green's functions and 
bare interaction vertices the ``high-energy parts'' of a diagram. 

%
\begin{minipage}{6.1in}
\medskip
\centerline{
\includegraphics[width=0.5\linewidth]{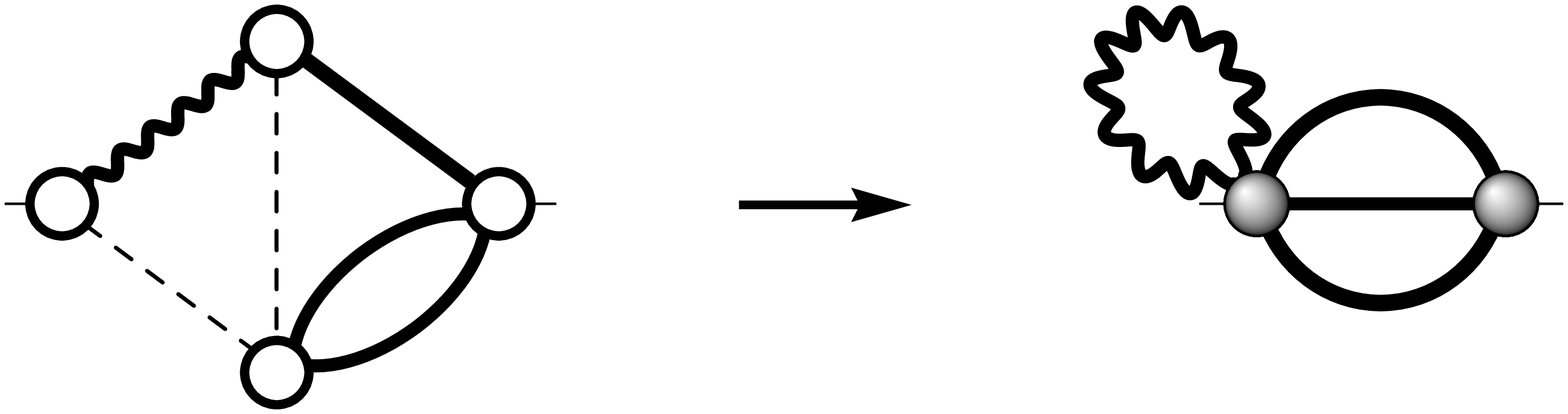}
}
\end{minipage}
\begin{quote}
\small
Fig. 1{\hskip 10pt}
Collapse of a diagram containing low-energy propagators (thick lines) 
and high-energy (dashed lines) propagators and bare vertices 
(open circles) into {\sl block} vertices (shaded circles) coupled to low-energy propagators.
\end{quote}
%
\medskip
\item The next step is to introduce self-energy diagrams. This
partial summation of diagrams is done in a way
that preserves the order of magnitude estimates. For
this purpose we merge {\sl connected} high-energy parts of a diagram
into  structureless `blobs', and obtain block diagrams consisting
of high-energy blobs and low-energy Green's 
functions. This procedure is shown  graphically in
 Figure 1.
 Diagrams which have  the same  block structures are of the
same order in {\small {\sl small}}. We sum them, and represent the sum
by the representative block diagram. These new self-energy diagrams
consist of low-energy Green's functions (smooth lines for electrons and
wiggly lines for phonons),  and high-energy vertices (shaded
 circles). 
The type of a vertex is characterized by the number
and type of its external links to low-energy Green's functions.  The
diagram in 
Figure 1,
  contains two high-energy vertices. One
is a two-phonon four-electron vertex, the other a four-electron vertex.
Next, we sum all self-energy insertions on the low-energy electron and
phonon Green's functions in the standard way to obtain the skeleton self-energy diagrams, with the bare low-energy Green's
functions replaced by exact low-energy Green's functions.

\item Up to this point we have not made any approximation.
The skeleton-diagram representation is still an exact formulation of
the many body problem. In order to proceed further we have
to make certain general assumptions. We assume that the construction of
skeleton diagrams by formally summing an infinite number of diagrams
preserves the original order of magnitude estimates. The elements of a
skeleton diagram, i.e. the full low-energy Green's functions and the
high-energy vertices should have the same order in {\small {\sl small}}
as their bare ancestors. Electron Green's functions are of order
{\small {\sl small$^{-1}$}}, while phonon Green's functions and high-energy
vertices are of order {\small {\sl small}}$^{\,0}$. We further assume
that the high-energy vertices vary with energy and momentum 
 on the high-energy scales, i.e. that  the summation
of high energy processes should not lead to new low-energy scales.  
With these assumptions we are able to classify each  skeleton
self-energy diagrams according to its order in  {\small {\sl small}}.
In addition to factors of {\small {\sl small}} coming from Green's
functions and vertices there are factors coming from phase-space
restrictions. The phase space for energy and momentum integrations is
restricted to the low-energy range, which leads to an additional 
 factor 
{\small {\sl small$^{\, \nu}$}}. The  power $\nu$ is always positive, and
depends on the topology of the low-energy lines and on the physical
dimensions of the system. 
\end{enumerate}

\subsection*{Order of Magnitude Estimates}

Figure 2 shows typical diagrams for the electronic self-energy together
with their order in {\small {\sl small}}. The given order of magnitude
refers to 3- and 2-dimensional metals. The phase space factors are
different for 1-dimensional metals, which leads to different order of
magnitude estimates, and to a break down of the Fermi-liquid model.
\par

Diagram (a) is the only one of order {\small {\sl small$^{\, 0}$}}; it
represents the effective band-structure potential seen by the
low-energy electrons. One obtains from this diagram  the quasiparticle
residue $a(\vec p_f)$, the Fermi surface, and the band-structure Fermi
velocities. The quasiparticle residue
is given by $(1-\partial \Sigma_0/\partial\epsilon)^{-1}$, where $\Sigma_0$ 
is the self-energy from   diagram (a).
We follow the traditional procedure  of Fermi-liquid theory
and  remove the quasiparticle residue from the low-energy propagators,
by combining it with  interaction vertices and self-energies. This leads to the
formation of renormalized interaction vertices and self-energies. Each link on a
vertex contributes a factor $\sqrt{a(\vec p)}$ to the renormalized
vertex; self-energies, $\epsilon\hat\tau_3$, and $\xi_0$ in Eqs. \ref{dyson}-\ref{dysonk}
 are multiplied by a factor $a(\vec p)$.  
In the following,  vertices and self-energies are  always understood as
renormalized quantities, so  the quasiparticle residue no longer appears 
in the theory. \par

The most important diagrams for the Fermi-liquid model are shown in the
second line of 
Figure 2.
They describe Landau's Fermi-liquid
interactions and the electronic pairing interactions (diagram (b)), the
leading order effects (Migdal's approximation) of the electron-phonon
coupling on electronic properties (diagram (c)), leading order effects
of impurities (the t-matrix diagrams (d1), (d2),...), and the external
perturbations (diagram (e)).  The third line of 
Figure 2 shows
diagrams  (of order {\small {\sl small}}) which
can safely be omitted.  Diagram (d1) describes  small changes of  Fermi
surface and Fermi velocities due to the alloying with impurities, and
diagram (f) describes the same kind of effects, induced by the small
changes in the effective band-structure potential due to quantum and
thermal smearing of the lattice positions. \par

\medskip
\centerline{
\includegraphics[width=0.75\linewidth]{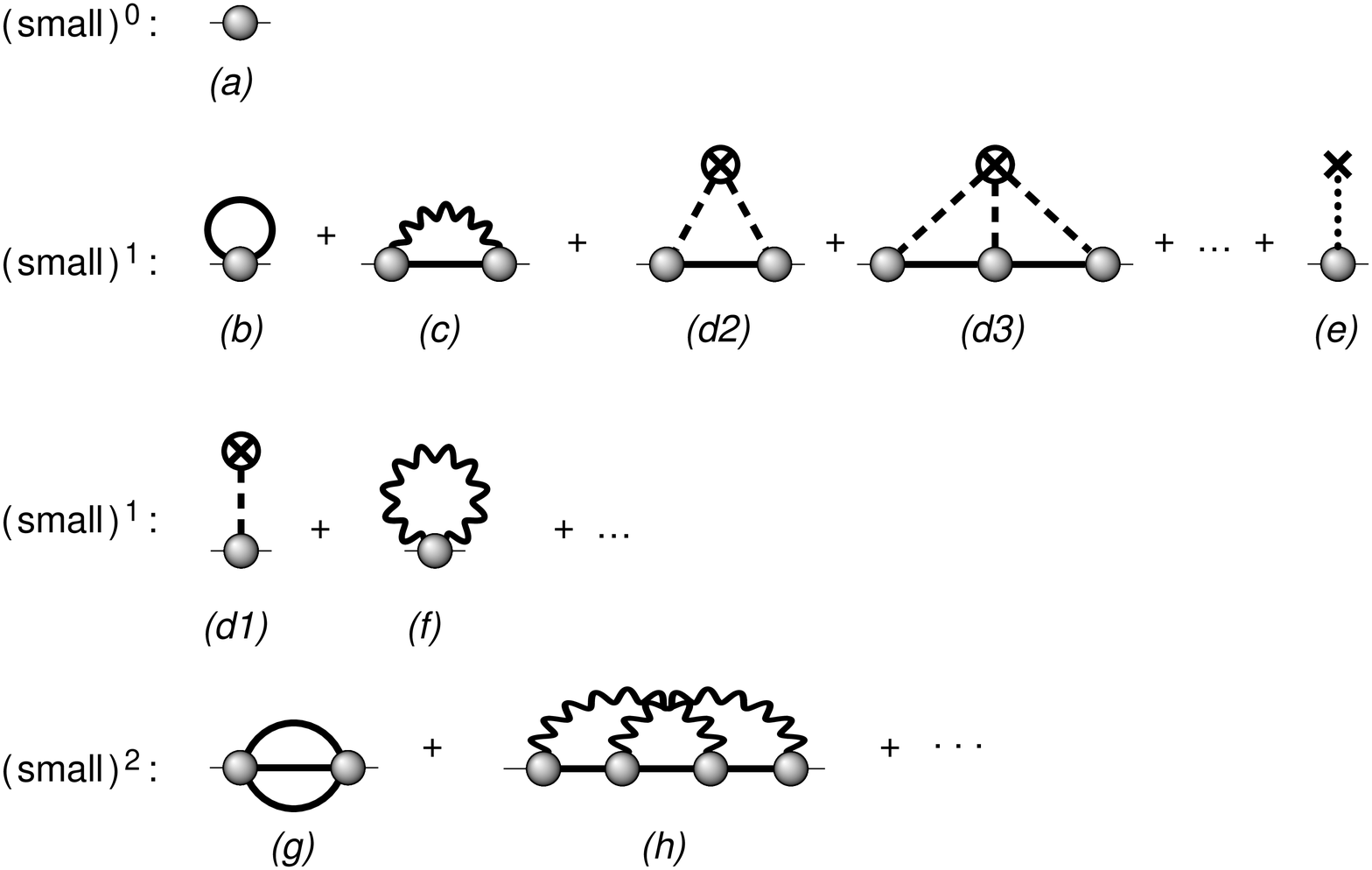}
}
\bigskip
\begin{quote}
\small
Fig. 2{\hskip 10pt}
Leading order electronic self-energy diagrams of Fermi-liquid theory. The block vertices (shaded
circles) represent the sum of all high-energy processes and give rise
to interactions between the quasiparticles (smooth propagator lines),
phonons (wiggly propagator lines) and impurities (dashed lines). The
order in the parameter `small'  is indicated for each diagram.
\end{quote}
\medskip

Two special diagrams of second order in {\small {\sl small}} are shown
in the fourth line. Diagram (g)  describes electron-electron
scattering. It leads to a contribution to the resistivity of a metal
$\propto T^2$. The same $T^2$ law also follows from diagram (h) which
represents a leading order correction to Migdal's theory. Both
contributions to the resistivity are of the same order of magnitude.
The Fermi-liquid model of metals comprises the effects of self-energy
processes described by the diagrams in the first two lines of 
Figure 2.
 Electron-electron scattering (diagram (g)) becomes
important in very clean metals with weak electron-phonon coupling, and
is often included in the Fermi-liquid model.  \par

The dominant contributions to the phonon self-energies are shown in
Figure 3. The  order of magnitude estimates refer to
phonons with wavelengths that are not too small, i.e. for $qa \ge$ {\small {\sl small}}, where $q$ is the phonon wave vector, and $a$ the lattice
constant.  The leading order effects are collected in diagram (a),
which gives the phonon dynamics in Born-Oppenheimer approximation. The
full vertex in diagram (a) contains the full effects of
electron-electron and electron-ion interactions. Hence, the phonon
dynamics obtained from diagram (a) is correct in the adiabatic limit.
Leading order non-adiabatic corrections are shown in the second line of
Figure  3. They are one order smaller than the adiabatic
diagrams.  Diagram (b) describes the damping of phonons due to the
coupling to electrons (Landau damping), and diagram (c) gives the
damping due to phonon-phonon coupling. Diagram (d) is included for
completeness. It describes small, anharmonic, T-dependent corrections
to the phonon frequencies, and can be neglected.

\medskip
\centerline{
\includegraphics[width=0.75\linewidth]{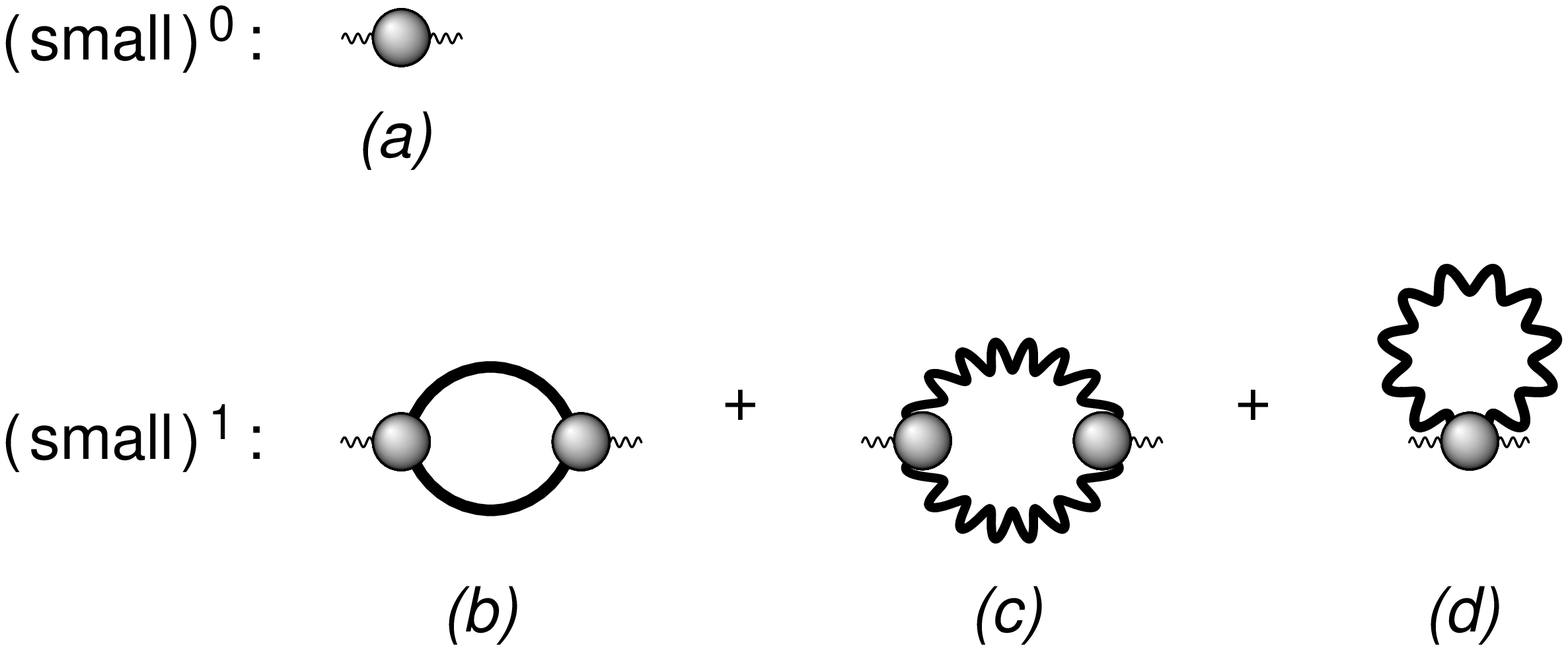}
}
\medskip
\begin{quote}
\small
Fig. 3{\hskip 10pt}
Leading order phonon self-energy diagrams of Fermi-liquid theory.
The notation is the same as that of Figure 2.
\end{quote}
%

\subsection*{R\'esum\'e}

We have reached our goal for this section, to identify and sum all
self-energy diagrams which contribute in leading orders in the
expansion parameters of the Fermi-liquid model - which includes a
strong electron-phonon coupling. If the premises of this expansion are
fulfilled and the expansion parameters are small, as it seems for
traditional metals, then we expect the Fermi-liquid model to give an
accurate description of the low-energy physics of metals in their
normal and superconducting states. On the other hand we paid a price
for  this accuracy. In forming block vertices we introduced new
interactions which replace the original Coulomb interaction and bare
electron-phonon interaction. These new interactions might have little
or no resemblence to the original interactions in strongly correlated
systems. Correlation effects may even reverse the sign of an
interaction, e.g. turn the repulsive Coulomb vertex into an attractive
block vertex for low-energy electronic excitations. The block vertices
are well defined and in principle measurable, but with present-day
many-body techniques we see no chance that they can be calculated
reliably from first principles. Thus, the Fermi-liquid theory of
correlated metals should be considered a phenomenological theory, with
Fermi-surface data and interaction vertices as material parameters to
be extracted from experiments. The superconducting transition
temperature depends on several of these material parameters, including
electronic interaction vertices and the electron-phonon vertex. Hence,
$T_c$ can be calculated (with a little bit of numerical effort) from
the  material parameters of the Fermi-liquid model. However,
understanding why the material parameters are optimized for  high
transition temperatures in some classes of materials but not in others
is not within reach of either the weak or the strong-coupling versions
of the Fermi-liquid model. Thus, the high accuracy of the
strong-coupling theory of superconductivity is of little help in the
search for new metals with higher T$_c$'s.

\section{Quasiclassical Theory of Strong-Coupling Superconductivity}

The Fermi-liquid model of strong-coupling superconductivity developed
in  section II is, in some sense, complete. Dyson's equations  together
with the self-energies (given in  diagramatic notation in Figs. 2-3,
form a closed set of equations which can be solved with the help of
computers.\footnote{ One might  drop the phase-space restrictions of
the Fermi-liquid model, and consider the Green's function lines in
Figs.2-3 as full Green's functions with low-energy and high-energy
parts.  Consequently, one has to replace Landau's interaction vertex
(diagram (b) of Fig. 2) and the other interaction vertices of the
Fermi-liquid model by new interactions, in order to avoid double
counting  of the high-energy Green's functions.  These models which
were   given collectively the name ``boson-exchange theory''
\cite{car90}, may be considered as generalized Fermi-liquid models.
They are often  discussed  as appropriate  models for high-T$_c$
superconductivity. The exchanged bosons can be  phonons,  paramagnons,
antiferromagnetic spin fluctuations, or others. We do not comment on
these models in this lecture but refer the interested disciples to the
recent literature.\cite{car90,rad92,mon93} This lecture is devoted to
the genuine Fermi-liquid model, which keeps consistently the leading
order effects in the expansion parameters of Fermi-liquid theory.}
Using fully quantum-mechanical equations such as  Dyson's equations in
the Fermi-liquid model is not wrong but also not optimal. By solving
Dyson's equations one calculates in a non-systematic way effects which
are beyond the  accuracy of the Fermi-liquid model. This generates an
unnecessary ``quantum-mechanical baggage'' which should be avoided,
especially if one is interested in solving more recent, and technically
more challenging problems of the theory of superconductivity.  The
appropriate  equation for the Fermi-liquid model which replaces Dyson's
equation, is a semiclassical transport equation, as first shown by
Landau for normal metals (normal Fermi liquids).  The conversion of
Dyson's equations for superconductors into transport equations was
first achieved by Eilenberger,\cite{eil68} Larkin and
Ovchinnikov,\cite{lar69} for superconductors in equilibrium, and
generalized by Eliashberg,\cite{eli72}  Larkin and
Ovchinnikov\cite{lar76} to non-equilibrium phenomena. The theory
established by these authors is called ``quasiclassical theory'' (for
reviews see \cite{ram86,lar86,eck81,ser83}).  It is the proper
generalization of Landau's  Fermi-liquid
theory of the normal state  to the superconducting state.  In the
following we sketch  the derivation of the quasiclassical transport
equations, give a brief interpretation, and write down the complete set
of equations of the Fermi-liquid model of strong-coupling
superconductors.  The theory presented in  this section holds for
general anisotropic superconductors with an arbitrarily complex
Fermi-surface, and general anisotropic interactions.  \par

\subsection*{$\xi$-Integration}

The one-electron  Green's functions ($\hat G^{R,A,K}$) depend on
momentum ($\vec p$), position ($\vec R$), energy ($\epsilon$), and time
($t$). This is more information than is required in order to specify
classical one-electron distribution functions. In classical systems the
energy is not an  independent variable since it is a prescribed
function of momentum and position.  Traditionally, classical
distribution functions depend on momentum, position, and time. A less
common,  but physically equivalent set of classical variables is
momentum direction, position, energy,  and time. The conversion from
one set to the other is a (non-canonical) coordinate transformation in
phase space.  The distribution functions of the Fermi-liquid model
depend on the latter variables, more precisely,  on the set $(\vec
p_f,\vec R;\epsilon,t)$, where $\vec p_f$ is a momentum vector  on the
Fermi surface.  The reduction from the set $(\vec p,\vec R;\epsilon,t)$
to $(\vec p_f,\vec R;\epsilon,t)$ is achieved by
``$\xi$-integration''.\cite{eil68,lar69} One obtains the
``quasiclassical propagators'' $\hat g^{R,A,K}(\vec p_f,\vec
R;\epsilon,t)$ by integrating out the non-classical variable of  the
low-energy Green's functions $\hat G^{R,A,K}_{l}(\vec p, \vec
R;\epsilon,t)$ in the following way:
\be\label{xiint}
\hat g^{R,A,K}(\vec p_f,\vec R;\epsilon,t)=\int d\xi(\vec p)\ {1\over a(\vec p)}\ \hat G^{R,A,K}
_{l}(\vec p, \vec R;\epsilon,t)\ .
\ee

The  ``bandstructure energy'', $\xi(\vec p)$, which is integrated out,
consists of   the kinetic energy $p^2/2m$ minus the chemical potential
plus  the effective band-structure potential defined by diagram (a) in
Fig. 1.  The band-structure energy vanishes at the Fermi surface, which
we characterize by the set of Fermi momenta $\vec p_f$.  Fermi-surface
integrals will be denoted by $\int d\vec p_f$. They are  normalized
such that 
\be\label{inte} \int{d^3p\over (2\pi)^3} \ ... \ = \ N_f\int
d\vec p_f\int d\xi(\vec p) \ ...  \ee $N_f$ is  the ``band-structure
density of states'' which is a material parameter of the Fermi-liquid
model. \par

The rules for evaluating the diagrams in Figs. 2-3 include
integrations over the momenta of internal Green's function lines. These
integrals can be transformed with Eq. \ref{inte} into $\xi$-integrals
and integrals over the Fermi surface.  The high-energy vertices depend
only weakly on $\xi$ for momenta near the Fermi surface. Thus, the
$\xi$-integrals act only on  the Green's functions,  and the
$\xi$-dependence of the vertices can be neglected in leading orders in
{\small {\sl small}}. The important consequence is that all electron
Green's functions in the self-energy  diagrams of the Fermi liquid
model are integrated over $\xi$,  and can be replaced by the
corresponding quasiclassical propagators. The factors $N_f$ are
absorbed into the vertices, which leads to dimensionless vertices and
interaction parameters. The   self-energies are now  functions of {\sl
quasiclassical propagators} and {\sl dimensionless interaction
vertices}.  The quasiclassical propagators describe physical properties
of low-energy electronic excitations with  momenta $\vec p$ near the
Fermi surface.  In the quasiclassical theory, however,  such
excitations are not described by their  momenta  but, equivalently, by
the momentum $\vec p_f$  nearest to $\vec p$, and the excitation energy
$\epsilon$.  The weak $\xi$-dependence of the self-energies can be
neglected in a leading order theory, such that  the self-energies
become  functions of $\vec p_f$ alone, in addition to  $\vec R$,
$\epsilon$, and $t$.  We use the symbols $\hat \sigma^{R,A,K}$ for the
self-energies on the Fermi surface in order to distinguish them from
the fully momentum dependent self-energies $\hat\Sigma^{R,A,K}$ of
section II. Similarly, the effective perturbations of diagram (e) in
Figure 2 can be taken at the Fermi surface, and will be denoted by $\hat
v(\vec p_f,\vec R;t)$.  \par

\subsection*{Quasiclassical Transport Equations}

Dyson's equations are formulated in section II in terms of the full
low-energy Green's functions. For a closed theory in terms of
quasiclassical propagators one has  to eliminate the low-energy Green's
functions in Dyson's equations  in favor of the quasiclassical
propagators.  This can be achieved by a procedure, traditionally used
for deriving transport equations from Dysons's equations (see
\cite{kad62}). One starts from  the two equivalent Dyson's equations
for $\hat G^{R,A}$,
\ber
& \left(\epsilon\hat\tau_3-\xi(\vec p)-\hat v(\vec p_f,\vec R;t)-
\hat\sigma^{R,A}(\vec p_f,\vec R;\epsilon,t)\right)\otimes a^{-1}(\vec p_f)\hat G^{R,A}_l
(\vec p,\vec R;\epsilon,t)=\hat 1 \ ,\label{ndyson1}\\
& a^{-1}(\vec p_f)\hat G^{R,A}_l(\vec p,\vec R;\epsilon,t)\otimes\left(\epsilon\hat\tau_3-\xi(\vec p)-\hat v(\vec p_f,\vec R;t)-
\hat\sigma^{R,A}(\vec p_f,\vec R;\epsilon,t)\right) =\hat 1 \ ,\label{ndyson2}
\eer
and for $\hat G^K$,
\ber
& \left(\epsilon\hat\tau_3-\xi(\vec p)-\hat v(\vec p_f,\vec R;t)-
\hat\sigma^{R}(\vec p_f,\vec R;\epsilon,t)\right)
\otimes a^{-1}(\vec p_f)\hat G^{K}_l
(\vec p,\vec R;\epsilon,t) 
-\nonumber\\
&\hat\sigma^{K}(\vec p_f,\vec R;\epsilon,t)\otimes a^{-1}(\vec p_f)\hat G^{A}_l
(\vec p,\vec R;\epsilon,t)=0\ ,\label{ndysonk1}\\
&a^{-1}(\vec p_f)\hat G^{K}_l(\vec p,\vec R;\epsilon,t) \otimes\left(\epsilon\hat\tau_3-\xi(\vec p)-\hat v(\vec p_f,\vec R;t)-
\hat\sigma^{A}(\vec p_f,\vec R;\epsilon,t)\right)
-\nonumber\\
& a^{-1}(\vec p_f)\hat G^{R}_l(\vec p,\vec R;\epsilon,t) \otimes
\hat\sigma^{K}(\vec p_f,\vec R;\epsilon,t)=0
\ .\label{ndysonk2}
\eer
The factors $a^{-1}$ in front of the Green's functions compensate the
factors $a$ included, by definition, in $\hat\sigma$, $\hat v$, etc.
(see section II).  One now  subtracts Eq. \ref{ndyson2} from
Eq. \ref{ndyson1}, and Eq. \ref{ndysonk2} from Eq. \ref{ndysonk1},
respectively, expands the operators $\xi(\vec p)$ in powers of the {\sl
small} spatial gradients,
\be\label{oper}
\xi(\vec p)\approx \xi \pm (i/2)\vec v_f\cdot\vec \nabla,
\ee
and $\xi$-integrates the resulting equations. The plus sign in Eq. \ref{oper} refers to a gradient acting to the right,
and the minus sign to one acting to the left. $\vec v_f=\nabla_p\xi$ is the Fermi velocity which depends on $\vec p_f$.

After subtraction the only remaining $\xi$-dependent terms are the
low-energy Green's functions which turn into the quasiclassical
propagators $\hat g^{R,A,K}$ by  $\xi$-integrating the equations. We
thus obtain the following transport equations for the quasiclassical
propagators
\be
\left[\epsilon\hat\tau_3-\hat v(\vec p_f,\vec R;t)-
\hat\sigma^{R,A}(\vec p_f,\vec R;\epsilon,t), \hat g^{R,A}(\vec p_f,\vec R;\epsilon,t)\right]_{\circ} 
+i\vec v_f\cdot\vec\nabla \hat g^{R,A}(\vec p_f,\vec R;\epsilon,t)= 0\ ,\label{transp1}
\ee
\ber\label{transp2}
& \left(\epsilon\hat\tau_3-\hat v(\vec p_f,\vec R;t)-
\hat\sigma^{R}(\vec p_f,\vec R;\epsilon,t)\right)\circ \hat g^{K}(\vec p_f,\vec R;\epsilon,t)
\\
&-\hat g^{K}(\vec p_f,\vec R;\epsilon,t)\circ  \left(\epsilon\hat\tau_3-\hat v(\vec p_f,\vec R;t)-
\hat\sigma^{A}(\vec p_f,\vec R;\epsilon,t)\right)\nonumber\\
&-\hat\sigma^{K}(\vec p_f,\vec R;\epsilon,t)\circ \hat g^{A}(\vec p_f,\vec R;\epsilon,t)+
 \hat g^{R}(\vec p,_f\vec R;\epsilon,t)\circ\hat\sigma^{K}(\vec p_f,\vec R;\epsilon,t)
\nonumber\\
&+i\vec v_f\cdot\vec\nabla \hat g^{K}(\vec p_f,\vec R;\epsilon,t)=0\ .\nonumber 
\eer
The $\circ$-product implies  here the following operation in the energy-time variables

\be\label{productqcl}
\begin{array}{l}
\hspace{-1.0em}
[\hat a\circ
\hat b](\vec p_f,\vec R;\epsilon,t)=
e^{{i\over 2}\left( 
\partial^a_{\epsilon}\partial^b_t
-\partial^a_{t}\partial^b_{\epsilon}\right)}
\hat a(\vec p_f,\vec R;\epsilon,t)\hat b(\vec p_f,\vec R;\epsilon,t)\,,
\end{array}
\ee
and the commutator $[\hat a,\hat b]_{\circ}$ stands for $\hat a\circ 
\hat b-\hat b\circ\hat a$. An important additional set of equations are  the
normalization conditions
\ber\label{norm}
&\hat g^{R,A}(\vec p_f,\vec R;\epsilon,t)
\circ \hat g^{R,A}(\vec p_f,\vec R;\epsilon,t)=\hat 1, \\
&\hat g^R(\vec p_f,\vec R;\epsilon,t)\circ \hat g^K(\vec p_f,\vec
R;\epsilon,t) + \hat g^K(\vec p_f,\vec R;\epsilon,t)\circ
\hat g^A(\vec p_f,\vec R;\epsilon,t)=0\ .\label{norma}
\eer
The normalization condition was first derived by
Eilenberger\cite{eil68} for superconductors in equilibrium. An
alternative, more physical derivation was given by
Shelankov\cite{she85}. The quasiclassical transport equations
(\ref{transp1},\ref{transp2}), supplemented by the normalization
conditions, Eqs. \ref{norm}-\ref{norma}, and the leading order self-energy diagrams
(Fig. 2(b-e)) are the fundamental equations of the Fermi-liquid theory
of superconductivity. The various transformations and simplifications
used to turn Dyson's equations into transport equations are consistent
with a systematic expansion to leading orders in {\small {\sl small}}.
\par

\subsection*{Quasiclassical Propagators}

The quasiclassical equations have a rich physical content. They cover
essentially all superconducting effects of interest ranging from
equilibrium phenomena, such as H-T phase diagrams and the structure of
vortex phases, to collective dynamics (e.g. Josephson effects, flux
motion, dynamics of phase-slip centers) and superconductivity far away
from equilibrium (e.g. stimulated superconductivity\cite{eli86}).  The
information about measurable quantities in equilibrium and
non-equilibrium is contained in the quasiclassical propagators, and
the transport equations are the tool for calculating these
propagators.  Because of the importance of the quasiclassical
propagators we  give a brief interpretation of their physical meaning,
and establish a convenient notation for their internal structure.  \par

The quasiclassical propagators are $4\times 4$-matrices whose
structure describes the quantum-mechanical internal degrees of freedom
of electrons and holes. The internal degrees of freedom are the spin
(s=1/2) and the particle-hole degree of freedom. The latter is of
fundamental importance for superconductivity.  In the normal state one
has an incoherent mixture of particle and  hole excitations, whereas
the superconducting state is  characterized by the existence of quantum
coherence between particles and holes.  This coherence is the origin of
persistent currents, Josephson effects, Andreev scattering, flux
quantization, and other non-classical superconducting effects.  The
quasiclassical propagators, in particular the combination $\hat
g^K-(\hat g^R-\hat g^A)$, are intimately related to the
quantum-mechanical density matrices which describe the
quantum-statistical state of the internal degrees of freedom.
Nonvanishing off-diagonal elements  in the particle-hole density matrix
indicate superconductivity, and the onset of non-vanishing off-diagonal
elements marks the superconducting transition. A convenient notation
for the matrix structure of the propagators (and similarly of the
self-energies) is
\begin{equation}\label{matrx}
\hat g^{R,A,K}=
\left(
\begin{array}{cc}
g^{R,A,K}+\vec g^{R,A,K}
\cdot\vec\sigma&\left(f^{R,A,K}
+\vec f^{R,A,K}\cdot\vec\sigma\right)i\sigma_y\\
i\sigma_y\left(\underline{f}^{R,A,K}+
\underline{\vec f}^{\raisebox{-1ex}{$\scriptstyle {R,A,K}$}}\cdot\vec\sigma
\right)&\underline{g}^{R,A,K}
-\sigma_y\underline{\vec g}^{R,A,K}
\cdot\vec \sigma\sigma_y
\end{array}\right)\ .
\end{equation}
The 16 matrix elements of $\hat g^{R,A,K}$ are expressed in terms of 4
spin-scalars ($g^{R,A,K}$, $\underline{g}^{R,A,K}$, $f^{R,A,K}$,
$\underline{f}^{R,A,K}$) and 4 spin-vectors ($\vec g^{R,A,K}$,
$\vec{\underline{g}}^{R,A,K}$, $\vec f^{R,A,K}$, $\underline{\vec
f}^{\raisebox{-1ex}{$\scriptstyle {R,A,K}$}}$). All matrix elements are
functions of $\vec p_f$, $\vec R$, $\epsilon$, and $t$.  The spin
scalars $g^{R,A,K}$, $\underline{g}^{R,A,K}$ carry the information on
the current density $\vec j(\vec R,t)$, the  charge density $n(\vec
R,t)$, the tunneling density of states $N_{tu}(\epsilon,\vec R;t)$, and
other spin-independent quantities. The current density is determined by
the scalar part of the Keldysh propagator, the Fermi velocity and the
density of states;
\be\label{current}
\vec j(\vec R,t)=2eN_f\int d\vec p_f\int{d\epsilon\over4\pi i} \vec v_f(\vec p_f)g^K(\vec p_f,\vec R;\epsilon,t).
\ee
The charge density in a metal is constant in  the low-frequency long-wavelength limit (local charge neutrality),
which leads to the condition,
\be\label{charge}
 2e^2N_f(1+A^s_0)\Phi(\vec R,t)+2eN_f\int d\vec p_f \int {d\epsilon\over 4\pi i} \left(1+
A^s_0(\vec p_f)\right)g^K(\vec p_f,\vec R;\epsilon,t)
=0\ .
\ee
The first term is the charge density fluctuation coming from
high-energy excitations under the influence of the electrochemical
potential $\Phi(\vec R,t)$. The compensating second term is the
charge-density fluctuation contributed  by the low-energy excitations.
The Fermi-liquid interaction parameters $A_0^s$ and $A^s_0(\vec p_f)$
are Fermi-surface averages of the spin-symmetric interaction vertex
$A^s(\vec p_f,\vec p_f^{\ \prime})$ of diagram (b) in Figure  2; $A^s_0
(\vec p_f)=\int d\vec p_f^{\ \prime} A^s(\vec p_f^{\ \prime},\vec p_f),
\ A^s_0 =\int d\vec p_f A^s_0(\vec p_f)$.  \par

The tunneling density of states is given by the usual formula of
Green's function theory,
\be\label{dos}
N_{tu}(\epsilon,\vec R;t)=N_f \int d\vec p_f 
\left[{-1\over \pi}{\cal I}m
\left( g^R(\vec p_f,\vec R;\epsilon,t)\right)\right]\ .
\ee

The spin vectors $\vec g^{R,A,K}$ in Eq. \ref{matrx} carry the information on the spin-magnetization, spin currents, 
and other spin-dependent properties. 
Finally, the off-diagonal terms $f$ and $\vec f$ are the anomalous propagators which are characteristic of the superconducting
state. A finite spin-singlet amplitude  $f^{K}$ indicates singlet pairing, and  a finite $\vec f^{K}$ 
implies  triplet pairing.
Finite pairing amplitudes can either arise spontaneously below $T_c$ or be induced by the contact to another superconductor 
(proximity effect). The anomalous propagators are not directly measurable, but couple via the transport equations to the
`measurable' propagators $g$, $\vec g$, and thus affect the observable properties,  often dramatically.
\par

\subsection*{Quasiclassical Self-Energies}

Finally, we discuss the analytical form for the self-energies of the
strong-coupling model, which are given in a compact diagramatic form by
the graphs in Figs. 2-3.  The original rules of Green's function theory
for assigning an analytical expression to each graph  can be readily
transformed into the rules of the quasiclassical theory.  The graphs in
both formulations are identical. In the quasiclassical scheme, however,
the thick lines stand for quasiclassical propagators, and the vertices
for renormalized interactions as discussed above. These interaction
vertices are  phenomenological quantities whose momentum dependences
must be in accordance with the symmetries of the system.  Momentum and
energy conservation holds at each vertex, and all electronic momenta
attached to a vertex are strictly confined to the Fermi surface. Thus,
integrals over internal electronic momenta are  2d Fermi-surface
integrals.  \par

The vertex in diagram (b) of Figure 2 represents  purely electronic
interactions in the particle-hole channel (Landau interactions,
$A^{s}(\vec p_f,\vec p_f^{\ \prime})$ and $A^{a}(\vec p_f,\vec
p_f^{\ \prime})$) and in the particle-particle channel
($\mu^{\ast}$-interactions, $\mu^{\ast s}(\vec p_f,\vec
p_f^{\ \prime})$ and $\mu^{\ast a}(\vec p_f,\vec p_f^{\ \prime})$). The
Landau interactions couple to the $g$-parts of the propagators, and the
$\mu^{\ast}$-interactions to  the $f$-parts. The superscripts $s$ ({\sl
symmetric, singlet}) refer to spin-independent interactions whereas $a$
({\sl antisymmetric}) and $t$ ({\sl triplet}) refer to  spin-spin
interactions such as exchange interactions ($A^a$) or triplet pairing
interactions ($\mu^{\ast t}$).  Below we give the explicit analytical
form of these self-energy terms, called ``mean-field'' ({\small {\sl
mf}}) self-energies, for systems with spin-rotation-invariant
interactions:
\ber
\label{mf}& &\hat \sigma_{sca,mf}^{R,A}(\vec p_f,\vec R;t)=\int {d\epsilon\over 4\pi i}
\int d\vec p_f^{\ \prime}\: A^s(\vec p_f,\vec p_f^{\ \prime})\, \hat g^K_{sca}
(\vec p_f^{\ \prime},\vec R;\epsilon,t), \\ \label{mf1}
& &\hat \sigma_{vec,mf}^{R,A}(\vec p_f,\vec R;t)=\int {d\epsilon\over 4\pi i}
\int d\vec p_f^{\ \prime}\: A^a(\vec p_f,\vec p_f^{\ \prime})\, \hat g^K_{vec}
(\vec p_f^{\ \prime},\vec R;\epsilon,t), \\ 
& &\hat \sigma_{sca,mf}^K=
\hat \sigma_{vec,mf}^K=0\ ,\label{mf2}
\eer

\ber\label{mustar}
& &\hat \Delta_{sca,mf}^{R,A}(\vec p_f,\vec R;t)=\int {d\epsilon\over 4\pi i}
\int d\vec p_f^{\ \prime}\: \mu^{\ast s}(\vec p_f,\vec p_f^{\ \prime})\, \hat f^K_{sca}
(\vec p_f^{\ \prime},\vec R;\epsilon,t),\\
\label{mustar1}& &\hat \Delta_{vec,mf}^{R,A}(\vec p_f,\vec R;t)=\int {d\epsilon\over 4\pi i}
\int d\vec p_f^{\ \prime}\: \mu^{\ast t}(\vec p_f,\vec p_f^{\ \prime})\, \hat f^K_{vec}
(\vec p_f^{\ \prime},\vec R;\epsilon,t),\\
\label{mustar2}& &\hat \Delta^K_{sca,mf}=
\hat \Delta^K_{vec,mf}=0\ .
\eer
We use   the following   short-hand 
notation for the various components of a $4\times 4$-propagator or self-energy
\ber\label{split}
&\hat g^{R,A,K} = \hat g_{sca}^{R,A,K}+\hat g_{vec}^{R,A,K}+\hat f_{sca}^{R,A,K}+\hat f_{vec}^{R,A,K},\\
\label{split1}& \hat \sigma^{R,A,K} = \hat \sigma_{sca}^{R,A,K}+\hat \sigma_{vec}^{R,A,K}+\hat\Delta_{sca}^{R,A,K}+
\hat \Delta_{vec}^{R,A,K},
\eer
where the matrices $\hat g_{...}$, $\hat \sigma_{...}$ contain the
diagonal parts in particle-hole space and $\hat f_{...}$, $\hat
\Delta_{...}$ the off-diagonal parts. The subscripts $sca$ and $vec$
refer to spin-scalar and spin-vector type matrix elements. The
mean-field self-energies are relatively simple compared to the
electron-phonon self-energies (denoted by the index $ ep$) shown
graphically in Figure 2(c).  The phonon-mediated interaction   is
retarded  and, as  a consequence, all three types of Keldysh
propagators $(R,A,K)$ enter the   self-energies, which  become
energy-dependent. The calculations of the energy dependences makes the
strong-coupling theory substantially more involved than the
weak-coupling theory,  where one approximates all interactions, including
the phonon-mediated interaction as instantaneous.  The self-energies
generated by Migdal's diagram Figure 2(c) have the form:
\ber\label{migdal1}
&\hat\sigma_{ep}^{R,A}(\vec p_f,\vec R;\epsilon,t)=
\int d\vec p_f^{\ \prime}
\int{d\omega\over 4\pi i}\biggl[\lambda^K(\vec p_f- \vec p_f^{\ \prime},\vec R;\omega,t)\hat g^{R,A}(\vec p_f^{\ \prime},\vec R;
\epsilon -\omega,t)+ \nonumber\\
&\lambda^{R,A}(\vec p_f, \vec p_f^{\ \prime},\vec R;\omega,t)\hat g^{K}(\vec p_f^{\ \prime},\vec R;
\epsilon -\omega,t)\biggr] ,
\eer
\ber\label{migdal2}
&\hat\sigma_{ep}^{K}(\vec p_f,\vec R;\epsilon,t)=\nonumber\\
&\int d\vec p_f^{\ \prime}
\int{d\omega\over 4\pi i}\biggl[\lambda^K(\vec p_f, \vec p_f^{\ \prime},\vec R;\omega,t)\hat g^{K}(\vec p_f^{\ \prime},\vec R;
\epsilon -\omega,t)- \nonumber\\
&\left(\lambda^{R}(\vec p_f- \vec p_f^{\ \prime},\vec R;\omega,t)- \lambda^{A}(\vec p_f- \vec p_f^{\ \prime},\vec R;\omega,t)\right)
\times\nonumber\\
&\left(\hat g^{R}(\vec p_f^{\ \prime},\vec R;
\epsilon -\omega,t)-\hat g^{A}(\vec p_f^{\ \prime},\vec R;\epsilon -\omega,t)\right)\biggr]\ . 
\eer
We have introduced a Bose propagator $\lambda^{R,A,K}(\vec p_f- \vec
p_f^{\ \prime} ,\vec R,\omega ,t)$ for the effective interaction
mediated by phonons.  This propagator is defined in terms of the
electron-phonon coupling $g_{\nu}(\vec p_f,\vec p_f^{\ \prime})$, and
the phonon propagators ${\cal D}_{\nu}^{R,A,K}(\vec q,\vec R;\omega,t)$,

\be\label{lambda}
\lambda^{R,A,K}(\vec p_f, \vec p_f^{\ \prime},\vec R,\omega ,t)=\sum\limits_{\nu}\mid g_{\nu}(\vec p_f,\vec p_f^{\ \prime})\mid^2
{\cal D}_{\nu}^{R,A,K}(\vec p_f- \vec p_f^{\ \prime},\vec R;\omega,t)\ .
\ee
The index $\nu$ sums over  the various branches of phonons,  including
acoustic and optical phonons.  \par

The above formulas hold for  electrons and phonons in  and out of
equilibrium. If the phonons are in equilibrium described by a phonon
temperature $T_{ph}$ one has
\be\label{a2qf}
\lambda^{R,A}(\vec p_f, \vec p_f^{\ \prime},\omega )=2\int\limits_0^{\infty}d\omega^{ \prime}{\alpha^2F(\vec p_f, \vec p_f^{\
\prime},\omega^{\prime})\, \omega^{\prime}\over \omega^{ \prime \ 2}-(\omega\pm i0)^2} \ 
\ee
and 
\be\label{pheq}
\lambda^{K}(\vec p_f, \vec p_f^{\ \prime},\omega )=\left(\lambda^{R}(\vec p_f, \vec p_f^{\ \prime},\omega )-
\lambda^{A}(\vec p_f, \vec
p_f^{\ \prime},\omega )\right)\coth\left({\omega\over k_BT_{ph}}\right)\, .
\ee
We introduced a momentum dependent spectral function $\alpha^2F(\vec
p_f, \vec p_f^{\ \prime},\omega)$ for the phonon-mediated interaction.
It is a generalization  of the ``Eliashberg function''
$\alpha^2F(\omega)$ to anisotropic interactions.  The averaged spectral
function
\be\label{a2qf1}
\alpha^2F(\omega)=\int d\vec p_f\int d\vec p_f^{\ \prime}\alpha^2F(\vec p_f, \vec p_f^{\ \prime},\omega)
\ee 
is measured in tunneling experiments on strong-coupling superconductors.
\par

The last set of self-energies of the Fermi-liquid model which we
discuss in this lecture is the impurity self-energy for randomly
distributed scattering centers.  One can neglect in leading order in
the ratio of Fermi wavelength and mean-free path ($1/(k_f\ell$)
coherent scattering of conduction electrons  (of wavelength $k_f^{-1}$)
from different impurities. This approximation leads to the t-matrix
diagrams shown in Figure 2.  The effects of impurities are described in
the Fermi-liquid model by an electron-impurity vertex, $v(\vec p_f,
\vec p_f^{\ \prime})$,  and the impurity concentration $c$.  The
impurity self-energy is given by
\be\label{imp}
\hat \sigma_{imp}^{R,A,K}(\vec p_f,\vec R;\epsilon,t) =c\ \hat t^{R,A,K} (\vec p_f,\vec p_f,\vec R;\epsilon,t),
\ee
where the single-impurity t-matrices are given by the solutions of the t-matrix equations
\ber\label{tmat1} 
&\hat t^{R,A}(\vec p_f,\vec p_f^{\ \prime},\vec R;\epsilon,t) = v(\vec p_f,\vec p_f^{\ \prime})+
\nonumber\\
& { \displaystyle \int d\vec p_f^{\ \prime\prime}\, v(\vec p_f,\vec p_f^{\ \prime\prime})
\, \hat g^{R,A}(\vec p_f^{\ \prime\prime},\vec R;\epsilon,t)
\circ \hat t^{R,A}(\vec p_f^{\ \prime\prime},\vec p_f^{\ \prime},\vec R;\epsilon,t)}
 \eer 
and
\be\label{tmat2}
 \hat t^K(\vec p_f,\vec p_f^{\ \prime},\vec R;\epsilon,t)=
 \int d\vec p_f^{\ \prime\prime}\, \hat t^R(\vec p_f,\vec p_f^{\ \prime\prime},\vec R;\epsilon,t)
\circ \hat g^K(\vec p_f^{\ \prime\prime},\vec R;\epsilon,t)
\circ \hat t^A(\vec p_f^{\ \prime\prime},\vec p_f^{\ \prime},\vec R;\epsilon,t).
\ee

\subsection*{R\'esum\'e}

We presented in this section the basic concepts and equations of the
Fermi-liquid theory of strong-coupling superconductors.  The important
equations are the quasiclassical transport equations 
(\ref{transp1}-\ref{transp2}) with the corresponding normalization conditions
 (Eqs. \ref{norm}-\ref{norma}) and the self-consistency equations for the self-energies
(\ref{mf}, \ref{mustar}, \ref{migdal1}, \ref{migdal2}, \ref{imp}).
These equations form a coupled system from which one calculates the
quasiclassical propagators $\hat g^{R,A,K}$. They may be interpreted as
quasiclassical density matrices, and they carry all the physical
information of interest.  A superconductor is specified in the
phenomenological Fermi-liquid model by a set of material parameters,
which include the Fermi surface data $\vec p_f$, $\vec v_f$, the purely
electronic coupling functions $A^{s,a}(\vec p_f,\vec p_f^{\ \prime})$
and $\mu^{\ast s,t}(\vec p_f,\vec p_f^{\ \prime})$, the phonon mediated
interaction $\lambda^{R,A}(\vec p_f, \vec p_f^{\ \prime},\omega)$, and
the impurity concentration $c$ and potential $v(\vec p_f,\vec
p_f^{\ \prime})$.  Calculations of physical properties of
strong-coupling superconductors follow the methods described above,
but generally require numerical work. We refer the interested
reader to original papers, review articles, and textbooks
\cite{all82,gin82,lar86,eli86,ram86,car90} for information on numerical
methods, results, and their interpretation.

\section{Linear Response}

The accuracy and power of the quasiclassical theory for strong-coupling
superconductors are based on the transport equations derived above and
the expansion of the self-energies in the expansion parameter {\small
{\sl small}}. In this part of the lecture we derive a compact solution
to the {\sl linearized} transport equations describing small deviations
from equilibrium.  We use Keldysh's method for non-equilibrium
phenomena, and concentrate on dynamical response functions, i.e. the
response of the superconductor to time-dependent external
perturbations, such as electromagnetic fields or ultrasound.  Static
response functions (the static spin-susceptibility, the superfluid
density, etc.) are more efficiently calculated using Matsubara Green's
functions.\cite{sca69,car90} \par

The derivation is general enough to cover isotropic and anisotropic
superconductors with strong electron-phonon, electron-electron and
electron-impurity interactions, and with a conventional or an
unconventional order parameter. The only important limitation of the
derivation is that it applies to the linear response of a
superconductor with a {\sl homogeneous} order parameter. This does not
exclude homogeneous current carrying states, but does exclude
inhomogeneous states, e.g. the response of superconductors in magnetic
fields in which vortices or superconducting-normal domains are
present.  Our goal is, again, to derive a response theory which is
exact to leading order in the expansion parameters of the Fermi-liquid
model. The linear response solutions  can be used to calculate a
variety of interesting transport properties. As a specific example, we
derive the electromagnetic response function for strong-coupling
superconductors with impurity scattering.  Early work on response
functions for strong-coupling superconductors  includes the work by
Ambegaokar and Tewordt\cite{amb64} on thermal conductivity and by Nam
\cite{nam67} on the electromagnetic response. The transport theory for
strong-coupling metals in the normal state was first derived in the
energy representation  by Prange and Kadanoff.\cite{pra64}  More
recent work is reviewed in Refs.\cite{car90} and \cite{gin92}.\par

We consider  the electrons and phonons to initially be in thermodynamic
equilibrium at temperature $T$. The metal is then subjected to a weak,
external perturbation, $\hat{v}(\vec{Q},\omega)
e^{i(\vec{Q}\cdot\vec{R}-\omega t)}$, of frequency $\omega$ and
wavevector $\vec{Q}$. For example, in the case of an electromagnetic
field
\be
\hat v(\vec Q,\omega)= -{e\over c}\vec v_f\cdot\vec A(\vec Q,\omega)\, ,
\ee
where $\vec A(\vec Q,\omega)$ is the Fourier component of the
electromagnetic vector potential. The goal is to solve the transport
equations (\ref{transp1})-(\ref{transp2}) to linear order in the
perturbation.  The first step is to linearize the transport equations
and normalization conditions in the perturbation $\hat v$, and the
deviations  induced by the perturbation,
\be
\delta\hat{g}^{R,A,K}(\vec{p}_f,\vec{R};\epsilon,t)=\hat{g}^{R,A,K}(\vec{p}_f,\vec{R};\epsilon,t)
- \hat{g}^{R,A,K}_o(\vec{p}_f;\epsilon)
\ee
and
\be
\delta\hat{\sigma}^{R,A,K}(\vec{p}_f,\vec{R};\epsilon,t)=\hat{\sigma}^{R,A,K}(\vec{p}_f,\vec{R};\epsilon,t)
- \hat{\sigma}^{R,A,K}_o(\vec{p}_f;\epsilon), 
\ee
where $\hat{g}_o^{R,A}(\vec{p}_f;\epsilon)$ and $\hat{\sigma}_o^{R,A}(\vec{p}_f;\epsilon)$
are the equilibrium propagators and self-energies. The equations for
the Fourier coefficients
$\delta\hat{g}^{R,A,K}(\vec{p}_f,\vec{Q};\epsilon,\omega)$,
$\delta\hat{\sigma}^{R,A,K}(\vec{p}_f,\vec{Q};\epsilon,\omega)$ further
simplify because all the remaining $\circ$-products reduce to standard
matrix products in $4\times 4$ spin$\times$particle-hole space.  \par

The $\circ$-products in the transport equations, which are defined by
\be
\hat{A}\circ \hat{B}\left({\epsilon ;t}\right) 
=e^{{{i}\over{2}}\left[{\partial^{A}_{\epsilon}\partial^{B}_{t}
-\partial^{A}_{t}\partial^{B}_{\epsilon}}\right]}\,
\hat{A}\left(\epsilon;t\right)
\hat{B}\left(\epsilon;t\right)
\,,
\ee
reduce to matrix products in the frequency representation if either
$\hat A$ or $\hat B$ is an equilibrium function. For example, 
if $\hat{A}(\epsilon;t)=\hat{A}_o(\epsilon)$ or $\hat{B}(\epsilon;t)=\hat{B}_o(\epsilon)$
then
\be
\hat{A}_o\left(\epsilon \right)\circ \hat{B}\left({\epsilon ;\omega}\right) 
=\hat{A}_o(\epsilon+{\omega\over 2}) \hat{B}\left({\epsilon ;\omega}\right)\,,
\ee
\be
\hat{A}\left({\epsilon ;\omega}\right)\circ \hat{B}_o\left(\epsilon \right)
=\hat{A}\left({\epsilon ;\omega}\right)\hat{B}_o(\epsilon-{\omega\over 2})\,,
\ee
where $\hat A(\epsilon;\omega)=\int dt \exp(i\omega t)\,\hat A(\epsilon;t)$, etc.
\par

The response of physical quantities such as  current density or
magnetization to a perturbation can be calculated directly from the
Keldysh response $\delta\hat{g}^K$. However, there is no transport
equation for $\delta\hat{g}^K$ alone; it is coupled directly to the
retarded and advanced functions, $\delta\hat{g}^{R,A}$, via the
transport equation (\ref{transp2}). On the other hand, the quasiclassical
equations for $\delta\hat{g}^{R,A}$ are closed and do not depend on
$\delta\hat{g}^K$. It is useful to decompose $\delta\hat{g}^K$ into a
``spectral response'' which involves $\delta\hat{g}^{R,A}$, and an
``anomalous response'' which is expressed in terms of the ``anomalous
propagator'' $\delta\hat{g}^{a}$.\cite{eli72,sch75} The spectral
response and the anomalous response follow from separate quasiclassical
transport equations. This separation is achieved by the decomposition,

\ber\label{defano}
&\delta\hat{g}^K(\vec{p}_f,\vec{Q};\epsilon,\omega)
=\tanh\left({\displaystyle {\epsilon-{\omega\over 2}\over 2T}}\right)
\delta\hat{g}^R(\vec{p}_f,\vec{Q};\epsilon,\omega)-
\tanh\left({\displaystyle{\epsilon+{{{\omega\over 2}}}\over
2T}}\right)\delta\hat{g}^A(\vec{p}_f,\vec{Q};\epsilon,\omega)\nonumber \\
 &\left(\tanh\left({\displaystyle{\epsilon+{{{\omega\over 2}}}\over
2T}}\right)-\tanh\left({\displaystyle{ \epsilon-{{{\omega\over 2}}}
\over 2T}}\right)\right)\delta \hat
g^a(\vec{p}_f,\vec{Q};\epsilon,\omega)\: .
\eer
Eq. \ref{defano} should be understood as the defining equation for
$\delta \hat g^a$.  In the following we calculate the spectral response
(the first two terms on the right side of Eq. \ref{defano}) and the
anomalous response (the last term) separately, and then combine them; only the sum, $\delta\hat{g}^{K}$, has a direct physical meaning.

\subsection*{Spectral Response}

First, consider the equations for retarded (R) and advanced (A)
propagators, $\delta\hat{g}^{R,A}$.  These solutions are important in
the superconducting state; they determine the linear response of the
quasiparticle excitation spectrum, and contribute  the spectral
response of the superconductor to the perturbing field.  The linearized
transport equations become,  
\ber\label{linra}
&{\displaystyle \left[
(\epsilon+{\omega\over 2})\hat{\tau_3} -\hat{\sigma}_o^{R,A}(\epsilon+{\omega\over 2})
\right]
\,\delta\hat{g}^{R,A}
- \delta\hat{g}^{R,A}\,
\left[
(\epsilon-{\omega\over 2})\hat{\tau_3}+\hat{\sigma}_o^{R,A}(\epsilon-{\omega\over 2})
\right]
}\nonumber\\ 
&{\displaystyle -\vec{Q}\cdot\vec{v}_f\,\delta\hat{g}^{R,A}
\quad
=\,
\left[\hat{v}+\delta\hat{\sigma}^{R,A}\right]
 \,\hat{g}_o^{R,A}(\epsilon-{\omega\over 2})
-
\hat{g}_o^{R,A}(\epsilon+{\omega\over 2})\,
\left[\hat{v}+\delta\hat{\sigma}^{R,A}\right]
\,,}
\eer
where the external field and the deviations of the propagators and
self-energies all have arguments $(\vec p_f,\vec Q;\epsilon,\omega)$.
Note that in addition to the explicit external field we have included
the deviations of the self-energies from equilibrium on the right-hand
side of the equation.  These deviations respresent what are called {\sl
vertex corrections} in a diagramatic representation of  linear response
theory in terms of 4-point Green's functions \cite{agd63,fet71,ric80}.
Such corrections describe the `dynamical screening' of the external
perturbation by the low-energy, long-wavelength excitations of the
correlated medium. The static (instantaneous) screening by
short-wavelength, high-energy correlations has already been included
through the renormalized vertex in diagram (e) of Fig. 2. Generally, vertex
corrections must be taken into account in order to satisfy
conservation laws, e.g. charge conservation. They guarantee, in
particular, that electron-impurity, electron-phonon, and
electron-electron collisions do not lead to a decay of conserved
quantities. Vertex corrections, i.e. self-consistently calculated
$\delta\hat \sigma$'s in the quasiclassical framework, also couple the
external perturbations to collective modes, which can lead to
collective resonances in the response functions. Investigations of
these collective mode contributions to the electromagnetic response go
back to the papers by Anderson,\cite{and58} Bogolyubov, et
al.\cite{bog58}, Tsuneto,\cite{tsu60}, Vaks, et al.\cite{vak61}, and
Bardasis and Schrieffer;\cite{bar61} and more recently in the context
of superconductors with non-s-wave order parameters.\cite{hir89,yip92}

We proceed by assuming that the equilibrium solutions have been
obtained by solving the equilibrium transport equations and self-energy
equations self-consistently.  For strong-coupling superconductors this
amounts to solving Eliashberg's equations numerically to obtain the
renormalized excitation energy
$\tilde{\epsilon}^{R,A}(\vec{p}_f,\epsilon)=\epsilon
 -\sigma_{ep}^{R,A}(\vec{p}_f,\epsilon) -\sigma_{imp}^{R,A}(\vec{p}_f,\epsilon)$ and renormalized order
parameter
$\tilde{\Delta}^{R,A}(\vec{p}_f,\epsilon)=\Delta_{mf}^{R,A}(\vec{p}_f)+\Delta_{ep}^{R,A}(\vec{p}_f,\epsilon)
+\Delta_{imp}^{R,A}(\vec{p}_f,\epsilon)$,
where $\sigma$ and $\Delta$ are the diagonal and off-diagonal upper
corners of the  equilibrium self-energy matrix, $\hat\sigma=(\epsilon -\tilde\epsilon)\hat\tau_3+\hat{\tilde\Delta}$. 
These renormalizations include, in general, Fermi-liquid interactions, electron-phonon interactions as well as 
magnetic and non-magnetic impurity scattering. Solving Eliashberg's equations amounts to a
self-consistent determination of the equilibrium self-energies $\hat\Delta^{R,A}_{sca,mf}$, $\hat \sigma^{R,A}_{ep}$,
and $\hat\sigma^{R,A}_{imp}$ from equations (\ref{mustar}, \ref{migdal1}, \ref{imp}).
In any case, we assume this step has been accomplished. The equilibrium retarded and advanced propagators are then given 
by\footnote{We restrict 
the derivation that follows to superconductors which are ``unitary'', i.e. {$\hat{\tilde{\Delta}}^2=-|\tilde{\Delta}|^2\,\hat 1$}. This
excludes non-unitary, spin-triplet order parameters that describe superconductors with spontaneous spin-ordering of the Cooper pairs. 
However, it is a minor limitation.}
\be\label{equilibrium}
\hat{g}_o^{R,A}(\vec{p}_f,\epsilon)=
{\tilde{\epsilon}(\vec{p}_f,\epsilon)\hat{\tau}_3
-\hat{\tilde{\Delta}}{}^{R,A}(\vec{p}_f,\epsilon)\over
d^{R,A}(\vec{p}_f,\epsilon)},
\ee
\be\label{denomi}
d^{R,A}(\vec{p}_f,\epsilon)=
-\frac{1}{\pi}\sqrt{
\tilde{\Delta}^{R,A}(\vec{p}_f,\epsilon)
\tilde{\Delta}^{R,A}(-\vec{p}_f,-\epsilon)^{\ast}
-\tilde{\epsilon}^{R,A}(\vec{p}_f,\epsilon)^2}
\,.
\ee
and are important inputs to the linear response theory.  The linearized
transport equations are supplemented by the normalization conditions
expanded through first order (we omit the Fermi momentum, external wave
vector and frequency unless they appear explicitly),
\be\label{normlin}
\left(\hat{g}_o^R\right)^2=\left(\hat{g}_o^A\right)^2=-\pi^2\hat 1
\qquad , \qquad
\hat{g}_o^{R,A}(\epsilon+{\omega\over 2})\,\delta\hat{g}^{R,A} +
\delta\hat{g}^{R,A}\,\hat{g}_o^{R,A}(\epsilon-{\omega\over 2}) = 0 \,.
\ee
We now make use of these normalization conditions to invert the matrix
transport equation. Special versions of this procedure have been used
to obtain the collisionless response for $\vec{Q}=0$
perturbations.\cite{ser83} Consider the transport equation for the
retarded function; the advanced function is simply obtained by
$R\rightarrow A$.  We use the equilibrium equation for $\hat{g}_o^R$,
$\left[
\epsilon\hat{\tau}_3 - \hat{\sigma}^{R} \, , \,
\hat{g}_o^{R}
\right] = 0$, 
and the normalization condition $(\hat{g}^R_o)^2=-\pi^2$ to replace
\be\label{repl}
\left[
(\epsilon\pm {\omega\over 2})\hat{\tau_3} -\hat{\sigma}_o^{R,A}(\epsilon\pm {\omega\over 2})
\right]
\longrightarrow
d^R(\epsilon\pm {\omega\over 2})\hat{g}_o^{R}(\epsilon\pm {\omega\over 2})
\,,
\ee
where $d^{R}$ is given by Eq. \ref{denomi}. We note that 
the impurity renormalizations drop out of the
equilibrium propagators $\hat g_0$ for isotropic superconductors (Anderson's
`theorem'). There is an exact cancellation of the contributions of impurities to the numerator and denominator 
of Eq. \ref{equilibrium}. Impurities 
 are clearly important in the transport properties.
Thus, for the replacement in Eq. \ref{repl} the denominator, $d^{R}$,
necessarily  includes all self-energy renomalizations  in the
definitions of $\tilde{\epsilon}^{R}$ and $\tilde{\Delta}^{R}$.\footnote{Self-energy terms proportional to the
unit matrix are not included in Eq. \ref{linra}, as they are rarely
important. For example, impurity scattering contributes a term
proportional to {$\hat{1}$} (see Eqs. \ref{tmat1}-\ref{tmat2}), but this
term is independent of energy, and thus, drops out of the linear
response equations as well.}

The normalization condition (Eq. \ref{normlin}) is used to pull
$\delta\hat{g}^{R}$ to the right of all matrices, so the linearized
transport equation becomes,
\ber\label{linr}
&{\displaystyle{\left[
d^R_+(\epsilon;\omega)
\hat{g}_o^{R}(\epsilon+{\displaystyle{{\omega\over 2}}}) - \vec{Q}\cdot\vec{v}_f
\right]
\,\delta\hat{g}^{R}=}}\nonumber\\
&{\displaystyle{
(\hat{v}+\delta\hat{\sigma}^{R})\,\hat{g}_o^{R}(\epsilon-{\displaystyle{{\omega\over 2}}}) -
\hat{g}_o^{R}(\epsilon+{\displaystyle{{\omega\over 2}}})\,(\hat{v}+\delta\hat{\sigma}^{R})
\,,}}
\eer
where 
\be\label{dplus}
d^{R,A}_{+}\equiv d^{R,A}(\epsilon+{\omega\over 2})+d^{R,A}(\epsilon-{\omega\over 2})
\,.
\ee
The matrix  acting on $\delta\hat{g}^{R}$ can be inverted using the normalization condition,
\be\label{invert}
\left[
d^{R}_{+}\hat{g}_o^{R}- \vec{Q}\cdot\vec{v}_f
\right]^{-1} =
\frac{-1}{(\pi d^{R}_{+})^2+(\vec{Q}\cdot\vec{v}_f)^2}
\left[
d^{R}_{+}\hat{g}_o^{R}+ \vec{Q}\cdot\vec{v}_f
\right]
\,.
\ee
Thus, the solutions for the retarded and advanced functions in terms of
the external perturbation plus vertex corrections become,
\ber\label{gRA}
&\delta\hat{g}^{R,A} =
{\displaystyle{{\left[
d^{R,A}_{+}\hat{g}_o^{R,A}(\epsilon+{\omega\over 2})+ \vec{Q}\cdot\vec{v}_f
\right]\over(\pi d^{R,A}_{+})^2+(\vec{Q}\cdot\vec{v}_f)^2}}
\times}\nonumber \\
&{\displaystyle{\left[
\hat{g}_o^{R,A}(\epsilon+{\omega\over 2})\,(\hat{v}+\delta\hat{\sigma}^{R,A}) -
(\hat{v}+\delta\hat{\sigma}^{R,A})\,\hat{g}_o^{R,A}(\epsilon-{\omega\over 2})
\right]\,.}}
\eer
Note that $\delta\hat{g}^{R,A}=0$ in the normal state;  the density of
states is unchanged by the external potential reflecting the
particle-hole symmetry that is built into the leading order
quasiclassical theory. However, in a superconductor an external field
does give rise to changes in the retarded and advanced propagators, and
the corresponding density of states for the Bogolyubov quasiparticles.

\subsection*{Anomalous and Keldysh Response}

In addition to  changes of   spectrum and wave functions of
quasiparticle states caused by the perturbation (spectral response),
the system also responds to the perturbation by changing the occupation
of these states. The external field can  accelerate thermally excited
quasiparticles, heat phonons by the driven motion of the quasiparticle
excitations, or produce new excitations by inducing transitions into
or out-of the condensate. Specific processes, which are not described
by the spectral response, are the excitation of quasiparticles for
frequencies above the pair-breaking edge in conventional weak-coupling
superconductors, or the excitation of collective modes of the order
parameter in unconventional superconductors.\cite{hir89,yip92}
All response processes not yet included in the spectral
response are called ``anomalous response'', following the terminology
introduced by Eliashberg\cite{eli72}. Thus, the physical ``Keldysh
response'' is the sum of the spectral and anomalous response.
\par

We start from the linearized transport equation for the Keldysh
propagator, in order to derive the transport equation for the anomalous
part $\delta \hat g^a$ of the response function $\delta\hat g^K$.
\ber\label{linkel}
&{\displaystyle\left[
(\epsilon+{\omega\over 2})\hat{\tau_3} -\hat{\sigma}_o^{R}(\epsilon+{\omega\over 2})
\right]
\,\delta\hat{g}^{K}
- \delta\hat{g}^{K}\,
\left[
(\epsilon-{\omega\over 2})\hat{\tau_3}+\hat{\sigma}_o^{A}(\epsilon-{\omega\over 2})
\right]\, 
 -}\nonumber\\
&
{\displaystyle{\hat{\sigma}_o^{K}(\epsilon+{\omega\over 2})
\,\delta\hat{g}^{A}\, 
- \delta\hat{g}^{R}\,
\hat{\sigma}_o^{K}(\epsilon-{\omega\over 2})
\, - 
\vec{Q}\cdot\vec{v}_f\,\delta\hat{g}^{K}
=\,}}\nonumber  \\
&{\displaystyle{\left[\hat{v}+\delta\hat{\sigma}^{R}\right]
 \,\hat{g}_o^{K}(\epsilon-{\omega\over 2})
+\delta\hat{\sigma}^{K}
 \,\hat{g}_o^{A}(\epsilon-{\omega\over 2})
\, -}}\nonumber\\
&{\displaystyle{\hat{g}_o^{K}(\epsilon+{\omega\over 2})\,
\left[\hat{v}+\delta\hat{\sigma}^{A}\right]
\, -\hat{g}_o^{R}(\epsilon+{\omega\over 2})\,
\delta\hat{\sigma}^{K}
\, .}}
\eer
We now use the equilibrium relation
\be\label{gkequ}
\hat{g}^K_0\,
=\, \tanh\left({\epsilon\over 2T}\right)
\, \left(\hat{g}^R_0-\hat{g}^A_0\right)\, ,
\ee
and Eq. \ref{defano} to eliminate $\hat g^K_0$ and $\delta\hat g^K$ in
Eq. \ref{linkel} in favour of $\hat g^{R,A}_0$, $\delta \hat g^{R,A}$,
and $\delta \hat g^a$.  Finally, we eliminate $\hat \sigma^K_0$  by
the  equilibrium relation
\be\label{selfeq}
\hat{\sigma}^K_0\,
=\, \tanh\left({\epsilon\over 2T}\right)
\, \left(\hat{\sigma}^R_0-\hat{\sigma}^A_0\right)\, ,
\ee
and $\delta\hat\sigma^K$ by the defining relation for $\delta\hat\sigma^a$, 
\ber\label{defanom}
&\delta{\sigma}^{K}
=\tanh{\displaystyle \left({\epsilon-{{\omega\over 2}}\over 2T}\right)}\,\delta\hat{\sigma}^{R}
-\tanh{\displaystyle \left({\epsilon+{{\omega\over 2}}\over 2T}\right)}\,\delta\hat{\sigma}^{A}
+\nonumber\\
&{\displaystyle \left(
\tanh\left({\epsilon+{{\omega\over 2}}\over 2T}\right)-
\tanh\left({\epsilon-{{\omega\over 2}}\over 2T}\right)
\right)}\,
\delta\hat{\sigma}^{a}
\, .
\eer
All terms with with factors $\tanh(...)$ cancel, and we obtain the
following equation for $\delta\hat g^a$
\ber\label{eqdela}
&{\displaystyle{\left[
d^R(\epsilon+{\omega\over 2})
\hat{g}_o^{R}(\epsilon+{\omega\over 2})\delta\hat{g}^a -
\delta\hat{g}^a\hat{g}_o^{A}(\epsilon-{\omega\over 2})
d^A(\epsilon-{\omega\over 2})
\right]
-\vec{Q}\cdot\vec{v}_f
\,\delta\hat{g}^{a}=}}
\nonumber\\
&{\displaystyle{
\left(\hat v+\delta\hat{\sigma}^{a}\right)\,\hat{g}_o^{A}(\epsilon-{\omega\over 2}) -
\hat{g}_o^{R}(\epsilon+{\omega\over 2})\,\left(\hat v+\delta\hat{\sigma}^{a}\right)
\,,}}
\eer

Eq. \ref{eqdela} can be solved 
with the help of the normalization condition for $\delta\hat g^a$,
\be\label{anom_norm}
\hat{g}_o^{R}(\epsilon+{\omega\over 2})\,
\delta\hat{g}^{a}+
\delta\hat{g}^{a}
\hat{g}_o^{A}(\epsilon-{\omega\over 2})\,=0\,,
\ee
as was done in the previous section for the spectral response;
$\hat \delta g^a$ is moved to right of all matrices using Eq. \ref{anom_norm}. Matrix inversion gives the result,
\be\label{gano}
\delta\hat{g}^{a} =
{{\displaystyle{\left[
d^{a}_{+}\hat{g}_o^{R}(\epsilon+{\omega\over 2})+ \vec{Q}\cdot\vec{v}_f
\right]}}\over{\displaystyle{(\pi d^{a}_{+})^2+(\vec{Q}\cdot\vec{v}_f)^2}}}
\left[
\hat{g}_o^{R}(\epsilon+{\omega\over 2})\,(\hat{v}+\delta\hat{\sigma}^{a}) -
(\hat{v}+\delta\hat{\sigma}^{a})\,\hat{g}_o^{A}(\epsilon-{\omega\over 2})
\right]\,.
\ee

We can now construct the explicit solution for the Keldysh response,
obtained by substituting $\delta\hat g^a$ from Eq. \ref{gano},
and $\delta\hat g^{R,A}$, from Eq. \ref{gRA}, into
Eq. \ref{defano}:
\ber\label{final}
&{\displaystyle{
\delta\hat{g}^K(\vec{p}_f,\vec{Q};\epsilon,\omega)\: =}}\nonumber\\
&\displaystyle{\tanh\left({\epsilon\mm{\omega\over 2}\over 2T}\right)
{\displaystyle{\left[
d^{R}_{+}\hat{g}_o^{R}(\epsilon\mp{\omega\over 2})+ \vec{Q}\mdot\vec{v}_f
\right]
\left[
\hat{g}_o^{R}(\epsilon\mp{\omega\over 2})(\hat{v}\mp\delta\hat{\sigma}^{R}) -
(\hat{v}\mp\delta\hat{\sigma}^{R})\hat{g}_o^{R}(\epsilon\mm{\omega\over 2})
\right] 
}\over\displaystyle{(\pi d^{R}_{+})^2+(\vec{Q}\mdot\vec{v}_f)^2}}}\nonumber\\
&{\displaystyle{-\tanh\left({\epsilon\mp{\omega\over 2}\over
2T}\right)
{\left[
d^{A}_{+}\hat{g}_o^{A}(\epsilon\mp\displaystyle{\omega\over 2})+ \vec{Q}\mdot\vec{v}_f
\right]
\left[
\hat{g}_o^{A}(\epsilon\mp\displaystyle{\omega\over 2})(\hat{v}\mp\delta\hat{\sigma}^{A}) -
(\hat{v}\mp\delta\hat{\sigma}^{A})\hat{g}_o^{A}(\epsilon\mm\displaystyle{\omega\over 2})
\right] 
\over (\pi d^{A}_{+})^2+(\vec{Q}\mdot\vec{v}_f)^2}}}
\nonumber\\
&{\displaystyle{+\left(\tanh\left({\epsilon\mp{\omega\over 2}\over
2T}\right)-\tanh\left({\epsilon\mm{\omega\over 2}\over 2T}\right)\right)\times
\nonumber}}\\
&{\displaystyle{{\left[
d^{a}_{+}\hat{g}_o^{R}(\epsilon\mp\displaystyle{\omega\over 2})+ \vec{Q}\mdot\vec{v}_f
\right]
\left[
\hat{g}_o^{R}(\epsilon\mp\displaystyle{\omega\over 2})\,(\hat{v}\mp\delta\hat{\sigma}^{a})-
(\hat{v}\mp\delta\hat{\sigma}^{a})\,\hat{g}_o^{A}(\epsilon\mm\displaystyle{\omega\over 2})
\right] 
\over (\pi d^{a}_{+})^2+(\vec{Q}\mdot\vec{v}_f)^2}
\ .}}
\eer

The solution for $\delta\hat g^K$ allows one to calculate any observable of
interest by straight forward integrations. The linear response of the electric
current to a perturbation $\hat v$, for instance, is given by
\be\label{deltaj}
\delta j(\vec Q,\omega)=2eN_f\int d\vec p_f\int{d\epsilon\over 4\pi i}\vec v_f(\vec p_f)
{1\over 4}Tr_4\left(\hat\tau_3\delta\hat g^K(\vec p_f,\vec Q;\epsilon,\omega)\right)
\,.
\ee

Eq. \ref{final} is as far as one can get in a calculation of the
linear response without having to specify the perturbation or
observable of interest. The general formula is a convenient starting
point for concrete calculations of response coefficients.  Several
cumbersome technical steps are already carried out in the derivation of
Eq. \ref{final}, and one can concentrate on a more difficult task of
response theory, which in most cases is, the calculation of the vertex
corrections $\delta\hat\sigma^{R,A,a}$. The consequences of
Fermi-surface anisotropy and anisotropic interaction parameters for the
dynamic response coefficients are also largely unexplored. Such
calculations require more specific knowledge of the material parameters of
the Fermi-liquid model.

\subsection*{Current Response}

As an application we derive formulas for the current response of an isotropic superconductor to an electromagnetic field. The external perturbation is
\be\label{pertur}
\hat{v}(\vec{p}_f;\vec{Q},\omega)=-\frac{e}{c}\vec{v}_f\cdot\vec{A}(\vec{Q},\omega)\hat{\tau}_3
\,,
\ee
where the vector potential describes both the field generated by
external sources and the field generated by induced currents in the
superconductor. The induced field is described in our formulation
through Maxwell's equation with the induced current as the source of
the induced field.\footnote{Short-range electromagnetic interactions
have been included in the renomalized electron-electron vertex.} In
order to evaluate the linear response formulas for
$\delta\hat{g}^{R,A,K}$ we must specify the relevant bandstructure
parameters, including the Fermi surface, $\vec{p}_f$, the normal-metal
density of states, $N_f$, the Fermi velocity $\vec{v}_f(\vec{p}_f)$, as
well as vertices for the electron-impurity interaction
$v(\vec{p}_f,\vec{p}_f{\,'})$, electron-electron interactions,
$\mu^{*s,a}(\vec{p}_f,\vec{p}_f{\,'})$ and $A^{s,a}(\vec{p}_f,\vec{p}_f{\,'})$,
and electron-phonon coupling, $g(\vec{p}_f,\vec{p}_f{\,'})$. The latter are
combined with the phonon propagators, ${\cal D}^{R,A,K}$ into
effective phonon interactions,
$\lambda^{R,A,K}(\vec{p}_f,\vec{p}_f{\,'},\omega)$. These input data,
combined with the equilibrium solutions for the propagators,
self-energies and distribution function are used to construct the
linear corrections to the propagators, $\delta\hat{g}^{R,A,K}$. Lastly,
we must evaluate the vertex corrections. In most cases this is a
difficult step because it requires a self-consistent determination of
the linear response propagators from the transport equation with the
vertex corrections given as functionals of $\delta\hat{g}^{R,A,K}$.
Nevertheless, the self-consistency step can be carried out
numerically.  In special cases the vertex corrections vanish, or they
can be calculated analytically.

Consider a conventional superconductor with strong
electron-phonon coupling and non-magnetic impurity scattering (which
also may be strong). We simplify matters by considering an isotropic
metal with isotropic impurity scattering. Thus, we assume an isotropic
Fermi surface; the bandstructure data that we require is the
total density of states at the Fermi surface, $N_f$, and the
(isotropic) Fermi velocity, $\vec{v}_f=v_f\,\hat{p}$. We also assume an
an isotropic electron-phonon interaction, $\lambda^{R,A,K}(\omega)$,
for equilibrium phonons. We further assume that the phonons remain in
equilibrium under the the action of the electromagnetic field, i.e. we
neglect any heating of the phonons by the electrical
current. Under these assumptions the vertex corrections vanish
for the current response to a {\sl transverse} electromagnetic field, 
i.e. $\vec{Q}\cdot\vec{A}=0$.\footnote{In principle, vertex corrections associated with excitations of the 
order parameter need to be included in a self-consistent calculation of the current response. For an isotropic system only the 
phase mode of the order parameter is excited by the electromagnetic field, 
and for a transverse field this vertex correction can be eliminated by a gauge transformation without 
violating charge conservation. However, for longitudinal fields a self-consistent calculation of the 
phase mode, and the corresponding vertex correction, are essential for enforcing the conservation of charge in the 
calculation of the longitudinal current response.} We thus can set 
$\delta\hat\sigma^R=\delta\hat\sigma^A=\delta\hat\sigma^a=0$
in Eq. \ref{final}.  

Having calculated the renormalized excitation energy,
$\tilde{\epsilon}^{R,A}(\epsilon)$, and gap,
$\tilde{\Delta}^{R,A}(\epsilon)$, by solving Eliashberg's equations
(see section III) we can construct the linear response of the transverse current
from Eqs.  
\ref{final}-\ref{deltaj}. We choose a real gauge,
$\tilde{\Delta}^{R,A}(\epsilon)^{*}=\tilde{\Delta}^{R,A}(-\epsilon)$,
in which case $\hat{\tilde{\Delta}}{}^{R,A}(\epsilon)
=\tilde{\Delta}^{R,A}(\epsilon)\,i\sigma_2\,\hat{\tau}_1$. For
isotropic, non-magnetic impurity scattering the impurity
renormalization drops out of the equilibrium propagators. The
excitation energy, $\tilde{\epsilon}^{R,A}(\epsilon)$, and gap
function, $\tilde{\Delta}^{R,A}(\epsilon)$, include, in the following,  only the
renormalizations due to the electron-phonon interaction, which we
assume has been calculated by solving Eliashberg's equations. The
impurity renormalization does contribute to the current response via
\ber
d^{R,A}_{+}(\epsilon)=-\frac{1}{\pi}
\left[
\sqrt{
\tilde{\Delta}^{R,A}(\epsilon\mp{\omega\over 2})^2
-
\tilde{\epsilon}^{R,A}(\epsilon\mp{\omega\over 2})^2
}
+
\sqrt{
\tilde{\Delta}^{R,A}(\epsilon\mm{\omega\over 2})^2
-
\tilde{\epsilon}^{R,A}(\epsilon\mm{\omega\over 2})^2
}
+\frac{1}{\tau}
\right]
\,,
\eer
\ber
&d^{a}_{+}(\epsilon)=-\frac{1}{\pi}
\left[
\sqrt{
\tilde{\Delta}^{R}(\epsilon\mp{\omega\over 2})^2
-
\tilde{\epsilon}^{R}(\epsilon\mp{\omega\over 2})^2
}
+
\sqrt{
\tilde{\Delta}^{A}(\epsilon\mm{\omega\over 2})^2
-
\tilde{\epsilon}^{A}(\epsilon\mm{\omega\over 2})^2
}
+\frac{1}{\tau}
\right]
\,,
\eer
where $1/\tau=$ is the impurity scattering rate, {\sl unrenormalized} by the electron-phonon interaction (similarly, the Fermi surface data, $v_f$ and $N_f$ refer to the bandstructure Fermi velocity and density of states).

Combining these results for $\tilde\epsilon^{R,A}$, $\tilde\Delta^{R,A}$, $d_+^{R,A,a}$ with Eqs. \ref{equilibrium}
and \ref{final} for $\hat g^{R,A}_0$ and $\delta\hat g^K$ one can calculate directly the   linear response function, 
$\delta\hat{g}^{K}$. Inserting it into Eq. \ref{deltaj} for the current, 
gives the current response to a transverse electromagnetic field,
\ber\label{conduct}
&\vec{j}(\vec{Q},\omega)=2\displaystyle
{e^2\over c} N_f 
\int\,d\vec{p}_f\,\vec{v}_f\,\vec{v}_f\cdot\vec{A}(\vec{q},\omega)\,
\int\,{\displaystyle{d\epsilon\over4\pi i}}\nonumber\\ 
&\times \Big\{
\tanh({\displaystyle{\epsilon\mm{\omega\over 2}\over 2T}})\:
{\displaystyle \frac{\displaystyle{\pi^2 d^{R}_+(\epsilon,\omega)}}
  {\displaystyle{\pi^2 d^{R}_+(\epsilon,\omega)^2 \mp (\vec{v}_f\cdot\vec{Q})^2}}
\left[
\frac{\tilde{\epsilon}^{R}_-\tilde{\epsilon}^{R}_+
     +\tilde{\Delta}^{R}_-\tilde{\Delta}^{R}_+}
     {\sqrt{(\tilde{\Delta}^{R}_-)^2
      \mm(\tilde{\epsilon}^{R}_-)^2}
      \sqrt{(\tilde{\Delta}^{R}_+)^2
      \mm(\tilde{\epsilon}^{R}_+)^2}}
+1
\right]\:+}\nonumber\\
&\displaystyle{
\tanh(\displaystyle{\epsilon\mp{\omega\over 2}\over2T})
\frac{\pi^2 d^{A}_+(\epsilon,\omega)}
  {\pi^2 d^{A}_+(\epsilon,\omega)^2 \mp (\vec{v}_f\cdot\vec{Q})^2}
\left[
\frac{\tilde{\epsilon}^{A}_-\tilde{\epsilon}^{A}_+
     +\tilde{\Delta}^{A}_-\tilde{\Delta}^{A}_+}
     {\sqrt{(\tilde{\Delta}^{A}_-)^2
      \mm(\tilde{\epsilon}^{A}_-)^2}
      \sqrt{(\tilde{\Delta}^{A}_+)^2
      \mm(\tilde{\epsilon}^{A}_+)^2}}
+1
\right]+}\nonumber\\
&\displaystyle{\left(\tanh(\displaystyle{\epsilon\mp{\omega\over 2}\over 2T}) 
- \tanh(\displaystyle{\epsilon\mm{\omega\over 2}\over 2T})\right)
\frac{\pi^2 d^{a}_+(\epsilon,\omega)}
  {\pi^2 d^{a}_+(\epsilon,\omega)^2 \mp (\vec{v}_f\cdot\vec{Q})^2}\:\times}\nonumber\\
&\displaystyle{\left[
\frac{\tilde{\epsilon}^{A}_-\tilde{\epsilon}^{R}_+
     +\tilde{\Delta}^{A}_-\tilde{\Delta}^{R}_+}
     {\sqrt{(\tilde{\Delta}^{A}_-)^2
      \mm(\tilde{\epsilon}^{A}_-)^2}
      \sqrt{(\tilde{\Delta}^{R}_+)^2
      \mm(\tilde{\epsilon}^{R}_+)^2}}
+1
\right]}
\Big\}\,.
\eer
The abbreviations $\tilde\Delta^{R,A}_{\pm}$ and $\tilde\epsilon^{R,A}_{\pm}$ stand for
$\tilde\Delta^{R,A}(\epsilon\pm \omega/2)$ and $\tilde\epsilon^{R,A}(\epsilon\pm \omega/2)$.
Eq. \ref{conduct} has the form $ j_i(\vec Q,\omega)=\sigma_{ij}
(\vec Q,\omega){i\omega\over c}A_j(\vec Q,\omega)=\sigma_{ij}(\vec
Q,\omega)
E_j(\vec Q,\omega)$,
 and one can read off directly the 
conductivity tensor $\sigma_{ij}(\vec Q,\omega)$. The above  result for the electromagnetic response is a basis for 
calculations of the optical conductivity or the surface impedance in isotropic strong-coupling
superconductors. It reduces to the result of Abrikosov, Gor'kov, and Khalatnikov (see 
\cite{agd63}) for weak coupling. The first studies of the
electromagnetic response for strong-coupling superconductors were those
of Nam\cite{nam67} who calculated the frequency-dependent conductivity of
Pb in the dirty limit, with the full frequency-dependent
strong-coupling gap $\tilde{\Delta}^R(\omega)$. Calculations of the
electromagnetic response of strong-coupling superconductors require numerical
work; investigations have been made  recently by  several groups \cite{lee89,dol90,shu91,nic91,dah92,nic92,lit92}. 
There is general agreement among different
groups when the same material parameters are used as input for the
calculations; one can be confident that calculations of the  electromagnetic response
in conventional, nearly isotropic strong-coupling 
superconductors are under control provided data on $\alpha^2F(\omega)$ and the Fermi-surface data are known.

\medskip
\centerline{
\includegraphics[width=0.75\linewidth]{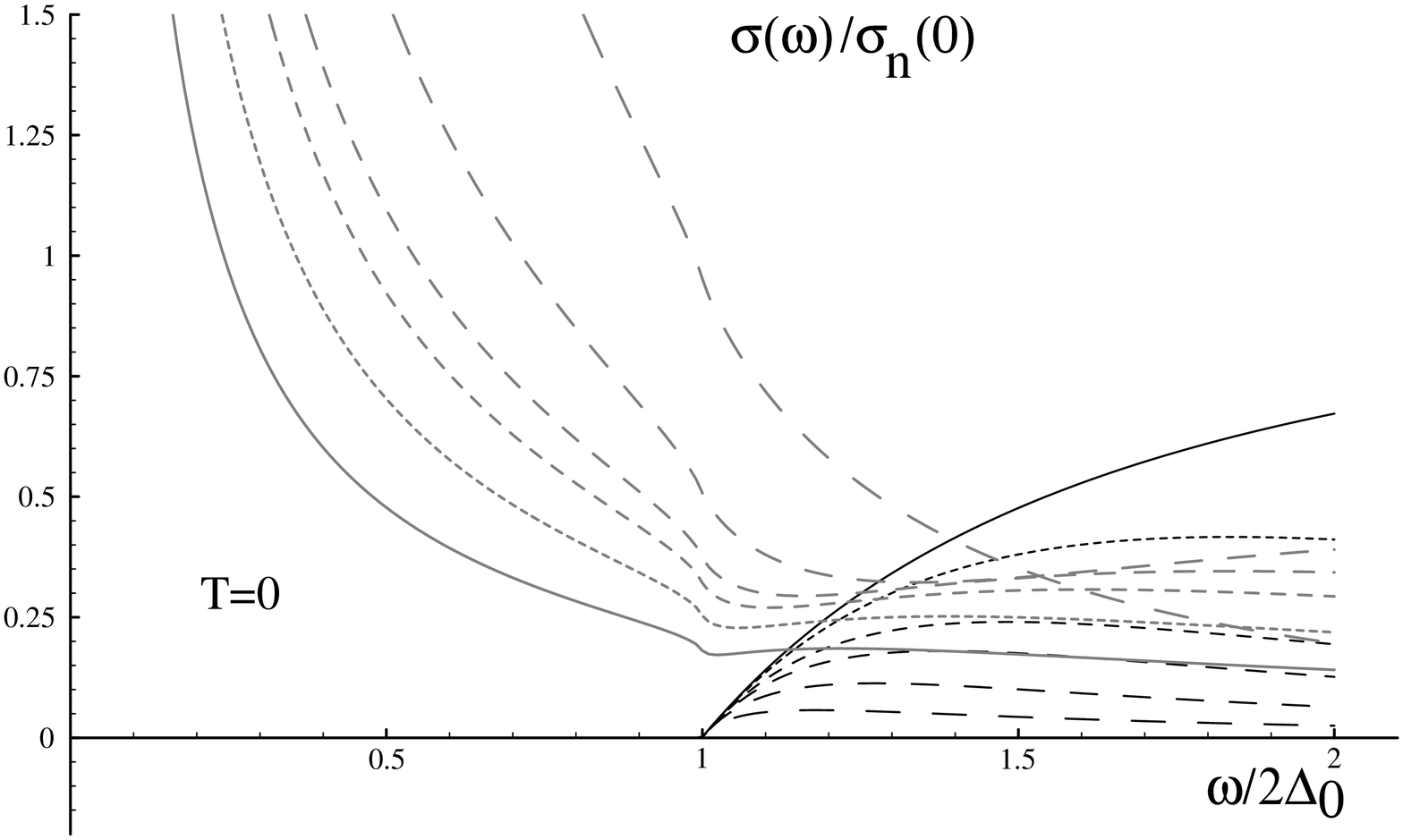}
}
\medskip
\begin{quote}
\small
Fig. 4{\hskip 10pt}
The long-wavelength conductivity for a weak-coupling superconductor at 
$T=0$. ${\cal R}e\,\sigma(\omega)$ curves are in black and ${\cal I}m\,\sigma(\omega)$ 
curves are in gray. The conductivity is normalized to the
d.c. conductivity of the normal state at $\omega=0$. The frequency is
scaled in units of twice the {\sl zero-temperature} gap, $2\Delta_0$. The
solid line corresponds the clean limit ($1/2\tau\Delta_0 =0$), while the
five dashed curves correspond to increasing levels of impurity scattering (decreasing 
$Re\,\sigma(\omega)$): $1/2\tau\Delta_0 = 0.3$, $0.5$, $0.75$, $1.0$, $2.0$.

\end{quote}
%
\medskip

As a specific example, we show results for the optical conductivity for several
models. Figures 4-6 are calculations of the conductivity in the
long-wavelength limit ($Q\rightarrow 0$) limit for a weak-coupling
superconductor with varying degrees of impurity scattering. Note the
onset of dissipation due to pair-breaking at $\omega=2\Delta(T)$. At
zero-temperature this is the only channel for dissipation (Fig. 4); at
higher temperatures the thermal excitations give rise to the Drude
conductivity (Fig. 5). The reactive component of the impedance, $\sim
{\cal I}m\sigma(\omega)$, is also shown in Fig. 4 for $T=0$. The low-frequency
limit, $\lim_{\omega\rightarrow 0}(\omega {\cal I}m
\sigma(\omega))$, determines the London penetration length.

\medskip
\centerline{
\includegraphics[width=0.75\linewidth]{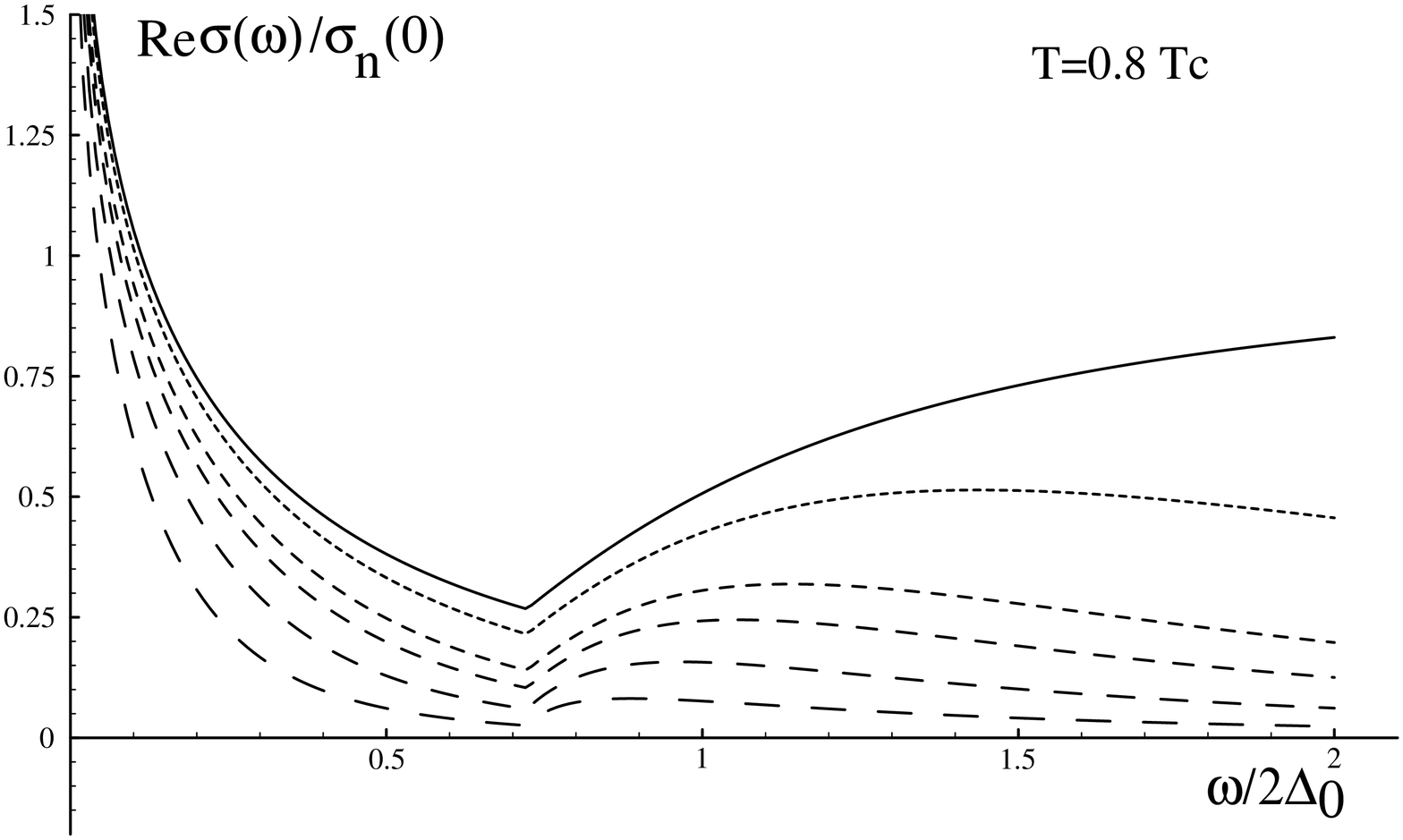}
}
\medskip
\begin{quote}
\small
Fig. 5{\hskip 10pt}
The real part of the long-wavelength conductivity for a weak-coupling superconductor at $T=0.8 T_c$. The notation is the same as that in Fig.
4 for the real part of the conductivity. Note the Drude contribution
to the conductivity associated with the thermally excited quasiparticles.
\end{quote}
%

%
\medskip

Calculations of the conductivity for strong-coupling superconductors
require more information on the material parameters of the metal, and
considerably more numerical computation. Figure 6 shows the real part
of the conductivity for an isotropic strong-coupling superconductor
with the electron-phonon coupling spectrum, $\alpha^2F(\omega)$, for
Pb$_{0.9}$Bi$_{0.1}$. The results are obtained from formula (\ref{conduct}), with the input data
$\tilde\epsilon$ and $\tilde\Delta$ taken from numerical solutions of
Eliashberg's equations by Allen \cite{all91}.  The long-wavelength
conductivity has been calculated for several values of the impurity
scattering scattering rate, $1/\tau$. In the clean limit the
conductivity rises steeply for $\omega/k_BT_c\simeq 10$, and shows
structure out to frequencies of order $\omega/k_BT_c\simeq 25$. The
steep rise in the conductivity at frequencies significantly above the
gap edge, $2\Delta(T)$, is due to the Holstein
effect.\cite{hol54,all71} In the superconducting state quasiparticles
that are produced by absorption of photons with frequencies above
$2\Delta$ can decay into phonon modes that are not otherwise 'infrared
active'. The anomalous absorption due to the Holstein effect starts at
a frequency $2\Delta/\hbar+\omega_p$, where $\omega_p$ is the frequency
of the emitted phonon.  For strong electron-phonon coupling and weak
impurity scattering these processes can dominate the threshold
absorption at $2\Delta$.\cite{lee89} With increasing impurity
scattering a structure at the gap edge ($2\Delta\approx 4k_BT_c$)
develops and becomes comparable in size to the Holstein absorption.
 This effect is shown in Fig. 6 for impurity scattering rates
increasing from $1/\tau T_c = 0.01$ (clean) to $1/\tau T_c = 20$
(dirty). In the extreme dirty limit ($1/\tau T_c = 100 >>
1/\tau_{phonon} T_c$) the phonon structure is essentially gone, and the
gap edge remains as the only significant structure.
  
\medskip
\centerline{
\includegraphics[width=0.75\linewidth]{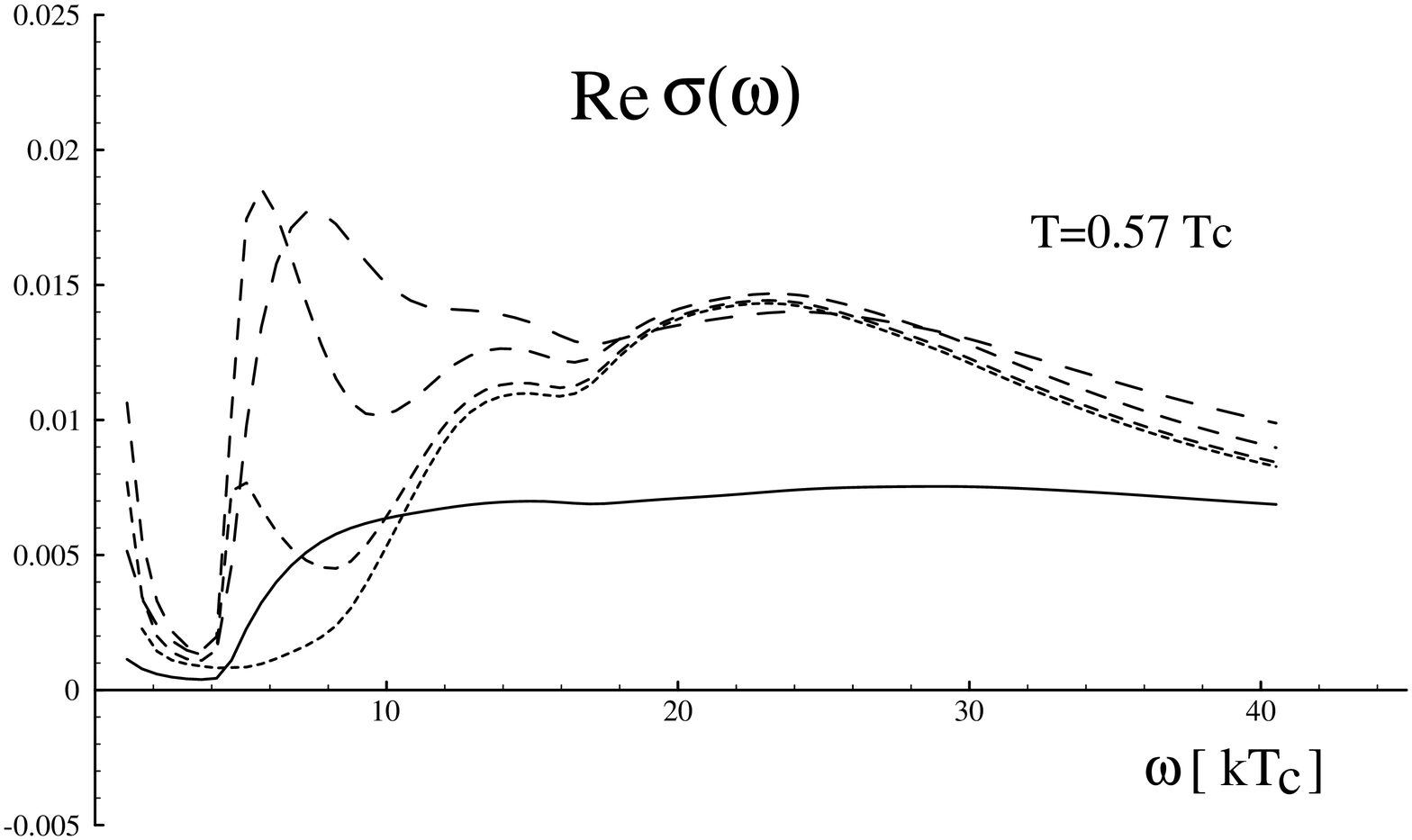}
}
\medskip
\begin{quote}
\small
Fig. 6{\hskip 10pt}
The real part of the long-wavelength conductivity at $T=0.57 T_c$ for a
strong-coupling superconductor with an electron-phonon spectrum,
$\alpha^2F(\omega)$, corresponding to the measured spectrum of
Pb$_{0.9}$Bi$_{0.1}$. The shortest dashed curve corresponds to the
clean limit ($1/\tau T_c =0.01$), while the solid curve corresponds to
the extreme dirty limit ($ 1/\tau T_c =100$). The remaining dashed
curves correspond to $1/\tau T_c =1.0$ (intermediate-clean: next
shortest dashing), $1/\tau T_c =5.0$ (intermediate-dirty: next to
longest dashing), and $1/\tau T_c =20.0$ (dirty: longest dashing).  The
x-axis is the frequency in units of $k_BT_c$, ($T_c =7.9\,K$ for
Pb$_{0.9}$Bi$_{0.1}$), and the y-axis is ${\cal R}e\,\sigma(\omega)$ in units
of ${2\over 3}N_fv^2_f e^2 /k_BT_c$, the Drude conductivity evaluated
with $1/\tau=k_BT_c$.

\end{quote}
%
\medskip
 
Fig. 6 gives a selected example of  the type of efects expected in the 
conductivity $\sigma(\omega)$ of strong coupling superconductors.
For detailed discussions of these effects, such as their temperature dependence
and their dependence on the electron-phonon coupling strength, we refer
to the original papers.\cite{lee89,dol90,shu91,nic91,dah92,nic92,lit92}

\bigskip
\section{conclusions}

These lectures present selected aspects of the largely established
theory of strong-coupling superconductivity. Given the increasing
scientific interest in superconductivity in strongly correlated metals
such as the cuprate high-T$_c$ superconductors we felt it important to
emphasize how electronic correlations are built into the
strong-coupling theory, in addition to the strong electron-phonon
coupling that is traditionally emphasized in textbooks on
strong-coupling superconductivity.\par

Our intent is also to give a reasonably complete collection of the
basic formulas of the strong-coupling theory, with a bit of explanation
and interpretation. Any evaluation of these formulas requires numerical
work. Hence, our selection of formulas is directed by the criteria of
generality and suitability for numerical work. In our opinion the most
efficient formulation of the strong-coupling theory is in terms of
quasiclassical transport equations, a method taken over from Landau's
theory of normal Fermi-liquids. Thus, a significant portion of the
lecture notes is devoted to the Fermi-liquid model of strong-coupling
superconductivity, whose central equations are the quasiclassical
transport equations. The Fermi-liquid model of strongly correlated
metals is a theory that straddles the border between microscopic and
phenomenological theories. The Fermi-liquid model is rooted in the
basic many-body theory of interacting electrons and ions, and is 
based on a small parameter expansion combined with a few general, but
plausible, assumptions about the relevant energy scales in metals.
However, the procedure of generating the Fermi-liquid model from its
microscopic antecedent introduces new, renormalized interactions (block
vertices) between the low-energy quasiparticles which cannot be
calculated with any confidence based on existing many-body algorithms.
Thus, the renormalized interactions that enter the Fermi-liquid model
are phenomenological parameters that must be determined from comparison
with experiments.
\par

The strong-coupling theory is  material oriented. In order to describe
important material effects we have kept the formulation of the
Fermi-liquid model general enough to cover  anisotropy, impurity
scattering and unconventional pairing.  One of the goals of the
strong-coupling theory was (and still is) to extract information about
the mechanisms of superconductivity from experimental data. In special
cases, such experimental data may contain valuable information, but in
general 
need sophisticated theoretical analysis for proper
interpretation. The most successful of the traditional methods is
quasiparticle tunneling combined with McMillan's strong-coupling
analysis. The tunneling current is one of many dynamical response
functions, and because of their potential importance we included a
section devoted to the linear response of superconductors. As an
example, we discuss the quasiclassical theory of the electromagnetic
response, and present selected results of strong-coupling
calculations.
\par

We have not discussed the strong-coupling model as a potential
mechanism for high-T$_c$ superconductivity. The reason is that it is
still an open and contraversial issue as to whether or not Fermi-liquid
theory is an appropriate model for the cuprates. 
Thus, the discussion of
the applicability of the strong-coupling model to the cuprates is
not yet settled and hence 
beyond the scope of these lecture notes.

\section{Acknowledgements}

\noindent The work of DR was supported by the  ``Graduiertenkolleg --   Materialien und Ph\"anomene bei sehr tiefen Temperaturen'' of the Deutsche Forschungsgemeinschaft. The work of JAS was supported by NSF grant no. DMR 91-20000 through the STC for Superconductivity, and the Nordic Institute for
Theoretical Physics in Copenhagen.

\end{document}